\documentclass[fleqn,usenatbib]{mnras}
\usepackage{newtxtext,newtxmath}
\usepackage[T1]{fontenc}

\usepackage{graphicx}	% Including figure files
\usepackage{bm,pdflscape}
\usepackage[encapsulated]{CJK}
\usepackage{ucs}
\usepackage[utf8x]{inputenc}
\usepackage[position=top]{subcaption}

\newcommand{\cntext}[1]{\begin{CJK}{UTF8}{bkai}#1\ignorespacesafterend\end{CJK}} % Typeset Chinese characters
\newcommand{\phn}{\phantom{0}} % phantom digit
\newcommand{\phd}{\phantom{.}} % phantom decimal point

\renewcommand{\vec}[1]{\bm{#1}} % vector
\newcommand{\diff}[1]{\mathrm{d}#1} % differential
\newcommand{\deriv}[2]{\diff{#1} / \diff{#2}} % derivative
\newcommand{\tder}[2]{\frac{\diff{#1}}{\diff{#2}}}
\newcommand{\va}[1]{\langle #1\rangle} % volume average
\newcommand{\sn}[2]{$#1\times10^{#2}$} % scientific notation

\newcommand{\taus}{\tau_\mathrm{s}} % dimensionless stopping time
\newcommand{\tausmin}{\tau_\mathrm{s,min}} % minimum dimensionless stopping time
\newcommand{\tausmax}{\tau_\mathrm{s,max}} % maximum dimensionless stopping time
\newcommand{\lambdac}{\lambda_\mathrm{c}} % critical wavelength
\newcommand{\Hg}{H_\mathrm{g}} % gas scale height
\newcommand{\Hp}{H_\mathrm{p}} % particle scale height
\newcommand{\Nsp}{N_\mathrm{sp}} % number of dust species
\newcommand{\mpk}[1]{m_\mathrm{p}^{(#1)}} % mass of the k-th dust particle
\newcommand{\rhog}{\rho_\mathrm{g}} % density of gas
\newcommand{\drhog}{\delta\rhog} % density fluctuation of gas
\newcommand{\np}{n_\mathrm{p}} % number density of particles
\newcommand{\rhop}{\rho_\mathrm{p}} % mass density of particles
\newcommand{\ug}{\vec{u}_\mathrm{g}} % gas velocity
\newcommand{\vpj}[1]{\vec{v}_{\mathrm{p},#1}} % dust velocity of species j
\newcommand{\vpk}[1]{\vec{v}_\mathrm{p}^{(#1)}} % velocity of k-th particle
\newcommand{\ugi}[1]{u_{\mathrm{g},#1}} % i-th component of gas velocity
\newcommand{\vpki}[2]{v_{\mathrm{p},#2}^{(#1)}} % i-th component of k-th particle velocity
\newcommand{\Dug}{\Delta\ug} % gas velocity deviation
\newcommand{\Dugi}[1]{\Delta\ugi{#1}} % i-th component of gas velocity deviation
\newcommand{\Dvpki}[2]{\Delta\vpki{#1}{#2}} % i-th component of k-th particle velocity deviation
\newcommand{\ugmi}[1]{\overline{\Dugi{#1}}} % i-th component of mean gas velocity deviation
\newcommand{\dugi}[1]{\delta\ugi{#1}} % component of gas velocity dispersion
\newcommand{\dvpi}[1]{\delta v_{\mathrm{p},#1}} % component of particle velocity dispersion
\newcommand{\Dpi}[1]{D_{\mathrm{p},#1}}
\newcommand{\Dpx}{\Dpi{x}}
\newcommand{\Dpz}{\Dpi{z}}
\newcommand{\cs}{c_\mathrm{s}} % speed of sound

\newcommand{\dda}{$\taus \in [10^{-3}, 0.1]$} % dust distribution with tausmax = 0.1
\newcommand{\ddb}{$\taus \in [10^{-3}, 2]$} % dust distribution with tausmax = 2

\title[Nonlinear Multi-species Streaming Instability]{%
    Streaming Instability with Multiple Dust Species:\\
    II.~Turbulence and Dust-Gas Dynamics at Nonlinear Saturation}

\author[C.-C.~Yang \& Z.~Zhu]{
Chao-Chin Yang (\cntext{楊朝欽})$^{1}$\thanks{E-mail: ccyang@unlv.edu (CCY)}
and Zhaohuan Zhu (\cntext{朱照寰})$^{1}$
\\
$^{1}$Department of Physics and Astronomy,
    University of Nevada, Las Vegas,
    4505 S.~Maryland Parkway, Box~454002,
    Las Vegas, NV~89154-4002, USA
}

\date{Accepted 2021 October 10. Received 2021 October 10; in original form 2021 August 10}
\pubyear{2021}

\begin{document}
\label{firstpage}
\pagerange{\pageref{firstpage}--\pageref{lastpage}}
\maketitle

\begin{abstract}
The streaming instability is a fundamental process that can drive dust-gas dynamics and ultimately planetesimal formation in protoplanetary discs.
As a linear instability, it has been shown that its growth with a distribution of dust sizes can be classified into two distinct regimes, fast- and slow-growth, depending on the dust-size distribution and the total dust-to-gas density ratio $\epsilon$.
Using numerical simulations of an unstratified disc, we bring three cases in different regimes into nonlinear saturation.
We find that the saturation states of the two fast-growth cases are similar to its single-species counterparts.
The one with maximum dimensionless stopping time $\tausmax=0.1$ and $\epsilon=2$ drives turbulent vertical dust-gas vortices, while the other with $\tausmax=2$ and $\epsilon=0.2$ leads to radial traffic jams and filamentary structures of dust particles.
The dust density distribution for the former is flat in low densities, while the one for the latter has a low-end cutoff.
By contrast, the one slow-growth case results in a virtually quiescent state.
Moreover, we find that in the fast-growth regime, significant dust segregation by size occurs, with large particles moving towards dense regions while small particles remain in the diffuse regions, and the mean radial drift of each dust species is appreciably altered from the (initial) drag-force equilibrium.
The former effect may skew the spectral index derived from multi-wavelength observations and change the initial size distribution of a pebble cloud for planetesimal formation.
The latter along with turbulent diffusion may influence the radial transport and mixing of solid materials in young protoplanetary discs.
\end{abstract}

\begin{keywords}
hydrodynamics
-- instabilities
-- methods: numerical
-- planets and satellites: formation
-- protoplanetary discs
-- turbulence
\end{keywords}

%-----------------------------------------------------------------------
\section{Introduction}

In our current understanding of the formation of planetesimals of $\sim$1--100\,km in size directly from mm/cm-sized dust particles in a protoplanetary disc, the streaming instability discovered by \cite{YG05} has been playing an important role \citep[see, e.g.,][and references therein]{BFJ16}.
When mono-disperse dust particles in a Keplerian disc interact with the surrounding gas via action-reaction pairs of drag forces, the system is linearly unstable.
In this regard, the disc should reach some type of turbulent saturation state driven by the streaming instability \citep{JY07}.
The turbulent diffusion of dust particles in such a state would lead to and maintain a vertically stratified dust layer \citep{YJ14,LYS18}, out of which planetesimals could form via gravitational collapse of local dust concentrations \citep{JM15,SA16,SYJ17,LYS19}.
However, it is known that to trigger strong clumping of solid material in such a disc for planetesimal formation requires a sufficiently high dust-to-gas ratio of \emph{column} densities $Z$ \citep{JYM09}.
Otherwise, a statistically steady state of the dust layer is maintained without any appreciable local dust concentrations.
This critical solid-to-gas ratio depends on the dust size \citep{CJD15,YJC17,LY21}, the background radial pressure gradient \citep{BS10b}, as well as the external turbulence \citep{YMJ18}.
Therefore, it appears that planetesimal formation involves the interplays between linear instability, nonlinear saturation, and triggers of strong clumping of solids.

\begin{table*}
\centering
\caption{Specifications of the simulation models.
	The columns are
	    (1)~model identifier,
		(2)~range of dimensionless stopping time,
		(3)~total solid-to-gas density ratio,
		(4)~growth regime of the instability,
		(5)~estimated critical wavelength,
		(6)~domain size,
		(7)~maximum simulation time, and
		(8)~maximum resolution.
    The lengths and times are in terms of the gas scale height $\Hg$ and the orbital period $P$, respectively.
	The power-law index of the dust size distribution is fixed at $q = -3.5$.
	The maximum number of discrete dust species we have investigated is $\max\Nsp = 64$.}
\label{T:specs}
\begin{tabular}{cccccccc}
	\hline
	Model & $\taus$ & $\epsilon$ & Regime & $\lambdac$ & $L_x = L_z$
	& $t_\mathrm{max}$ & Max.\ Resolution\\
	            &         &            &        & ($\Hg$)    & ($\Hg$)     & ($P$)\\
	(1)   & (2)     & (3)        & (4)    & (5)        & (6)
	& (7)              & (8)\\
	\hline
	Af & $[10^{-3}, 0.1]$ & 2   & fast & 0.0052       & 0.04
	& \phn\phn50       & $512\times512$\\
	As & $[10^{-3}, 0.1]$ & 0.2 & slow & 0.031\phn    & 0.2\phn
	&       5000       & $256\times256$\\
	B  & $[10^{-3}, 2]$   & 0.2 & fast & 0.31\phn\phn & 2\phd\phn\phn
	&    \phn200       & $512\times512$\\
	\hline
\end{tabular}
\end{table*}

\begin{figure*}
	\includegraphics[width=0.9\textwidth]{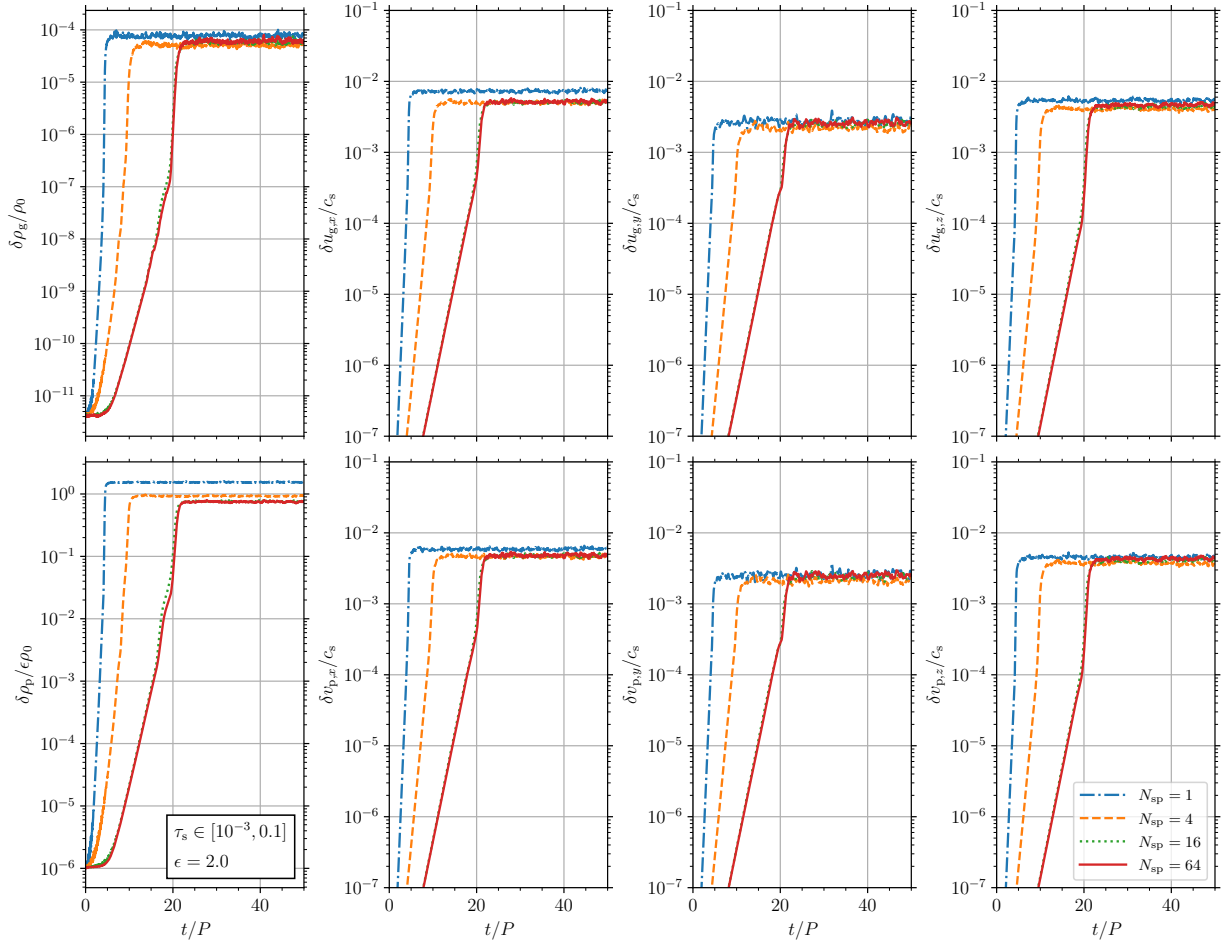}
	\caption{Dispersions as a function of time for Model~Af.
		The top and the bottom rows are for the gas and for the dust particles, respectively.
		The columns from left to right show the densities and the three components of the velocities.
		Different lines represent systems with different number of discrete dust species $\Nsp$ from one to 64.
		The densities are normalised by the mean densities, while the velocities are normalised by the speed of sound $\cs$.}
	\label{F:disp1}
\end{figure*}

\begin{figure*}
	\includegraphics[width=0.9\textwidth]{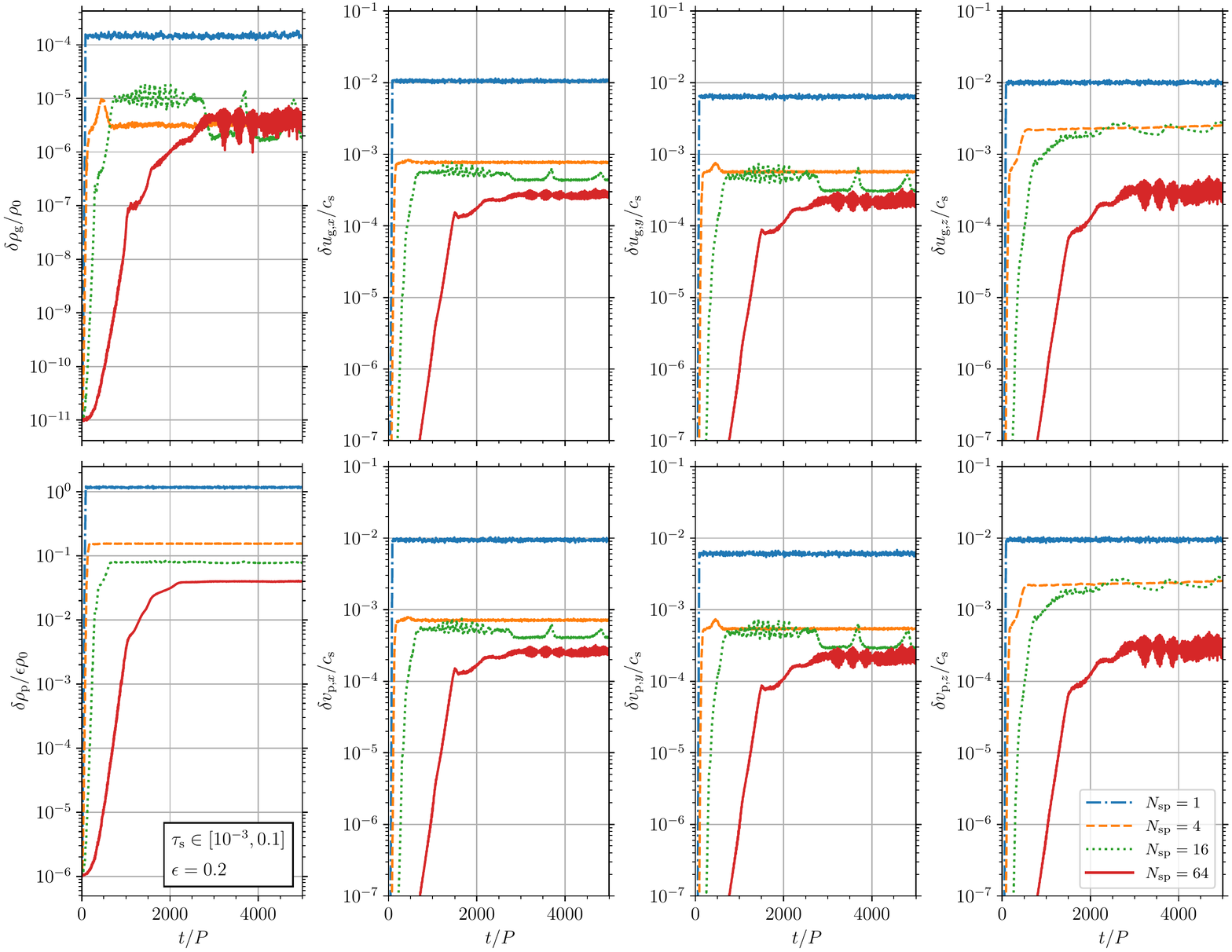}
	\caption{Similar to Fig.~\ref{F:disp1} except for Model~As.}
	\label{F:disp2}
\end{figure*}

\begin{figure*}
	\includegraphics[width=0.9\textwidth]{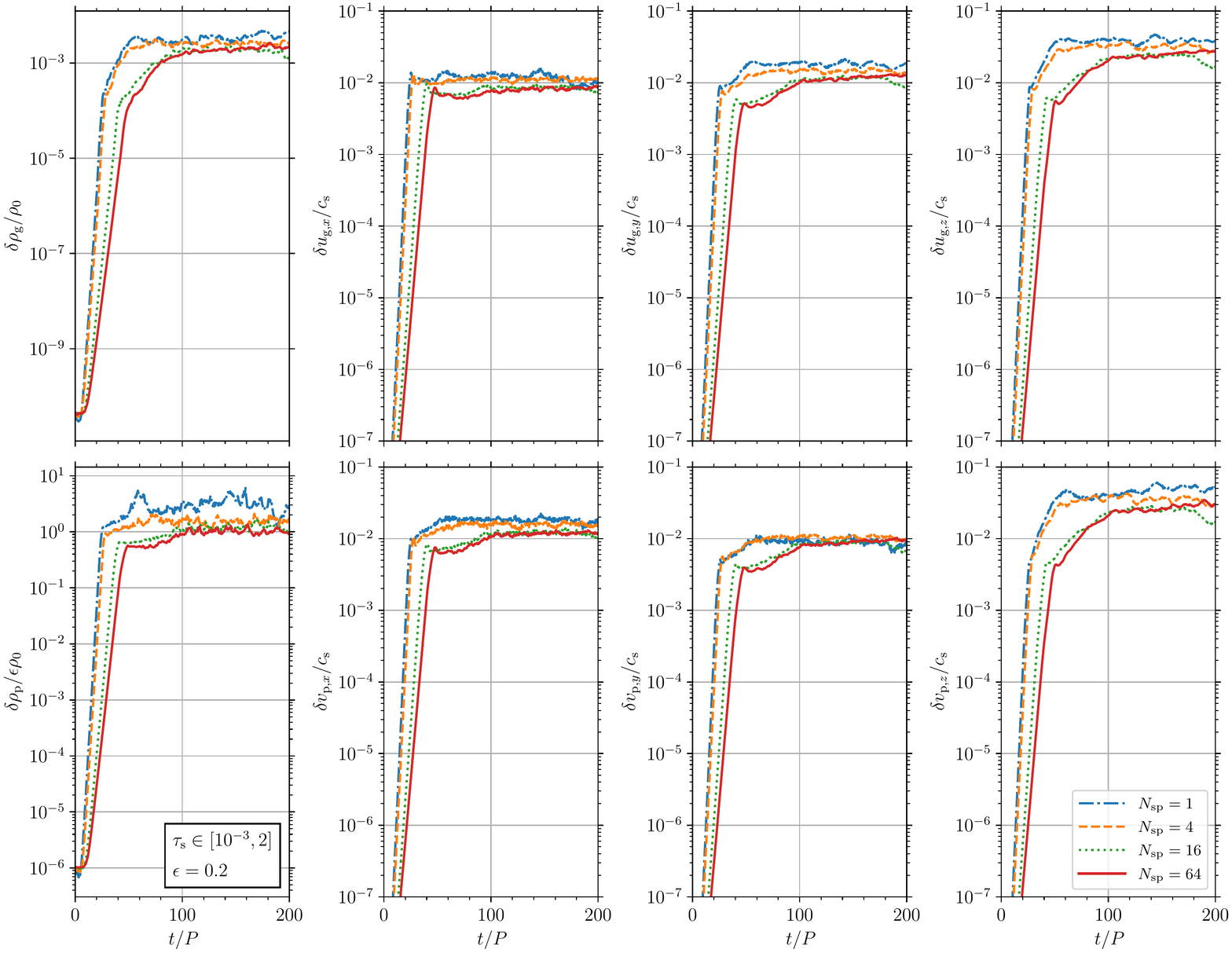}
	\caption{Similar to Fig.~\ref{F:disp1} except for Model~B.}
	\label{F:disp3}
\end{figure*}

In contrast to the previous works on the streaming instability with mono-disperse dust species, recent linear analyses of the instability with a distribution of dust sizes 
showed a much more complicated picture.
\cite{KB19} found that for some dust-size distributions, the growth rate of the instability decreases monotonically with increasing number of discrete dust species that represent the distribution, and it did not appear to approach to some finite value.
In \citet[and hereafter Paper~I]{ZY21}, we extended their linear analysis by conducting a systematic parameter study of much larger space, and determined that there exists two distinct regimes for the instability, separated by a sharp boundary.
When the largest particles in the distribution have a dimensionless stopping time of $\tausmax \gtrsim 1$ \emph{or} the total solid-to-gas density ratio $\epsilon \gtrsim 1$, the linear growth is fast, comparable to or faster than the orbital timescale, and the convergence could be achieved by a small number of dust species.
On the contrary, when $\tausmax \lesssim 1$ \emph{and} $\epsilon \lesssim 1$, the linear growth is much slower, and it may require a huge number of species to reach convergence of the growth rate, if any.
Similar conclusions were found when taking the dust-size distribution to the continuous limit \citep{PML20,MLP21}.
Interestingly, strong clumping of solids with a distribution of sizes could still be triggered in a vertically stratified disc, as long as $Z$ is sufficiently high \citep{BS10c,SJL21}.

To gain more insight into how a dust layer with a distribution of sizes is established with the streaming instability and in turn how strong clumping can be triggered, it is essential to study the intermediate stage, i.e., the nonlinear saturation of the streaming instability in an unstratified disc.
Without the complication introduced by the vertical gravity of the central star, one could observe how the saturation state driven by the instability itself is reached and more importantly, the statistical properties of the turbulence, if any, including velocity dispersion and turbulent stress of the gas, dust radial drift and diffusion, and density and size distribution of the dust, among others.
These pieces of information could be used as a guide and predict the properties near the mid-plane of a vertically stratified disc, which helps isolate the effects of vertical sedimentation of the gas and the dust particles by comparison.
We note that \cite{LY21} recently found no apparent connection between the linear growth of the streaming instability in an unstratified disc and the strong clumping of solids in a vertically stratified one.
Therefore, we believe it is even more imperative to further investigate the nonlinear saturation of the streaming instability before understanding its connection to vertically stratified discs.

In this work, we present the first systematic study of the nonlinear saturation of the streaming instability with multiple discrete dust species in an unstratified disc, using numerical simulations.
Our simulation setup and model parameters are described in Section~\ref{S:method}.
In Section~\ref{S:sat}, we observe the simulations reaching the saturation state and report the properties of the resulting gas turbulence.
We discuss at length in Section~\ref{S:dust} the dust dynamics at the saturation state, including density distribution, kinematics, and turbulent diffusion, along with implications for vertical sedimentation and radial transport of the dust particles.
Finally, we conclude in Section~\ref{S:cr} with potential consequences of our results on observations, radial mixing of solids, and planetesimal formation.

%-----------------------------------------------------------------------
\section{Methodology} \label{S:method}

Building upon the findings in Paper~I, we focus our attention on three distinct cases: (1)~\dda, $\epsilon = 2$, (2)~\dda, $\epsilon = 0.2$, and (3)~\ddb, $\epsilon = 0.2$, where $\taus$ and $\epsilon$ are the dimensionless stopping time of the dust particles and the total solid-to-gas density ratio, respectively.
The power-law index of the dust size distribution is fixed at $q \equiv \deriv{\ln \np}{\ln\taus} = -3.5$, where $\diff{\np(\taus)}$ is the number density of dust particles with stopping times between $\taus$ and $\taus + \diff{\taus}$.
As shown by fig.~2 of Paper~I, these cases are in the fast-, slow-, and fast-growth regimes, respectively.
Moreover, table~1 of Paper~I listed for each case a mode with dimensionless wave number $K = K_x = K_z$ that has approximately the maximum growth rate and is located near the ``knee'' of similarly fast growing modes in the Fourier space (see, e.g., fig.~8 and fig.~9 there).
With this in consideration, we adopt a computational domain $L_x$ by $L_z$ that can accommodate several critical wavelengths $\lambdac \simeq (2\pi\Pi / K)\Hg$, where $\Pi = 0.05$ is the dimensionless radial pressure support and $\Hg$ is the scale height of the gas.
These three cases are specified in Table~\ref{T:specs}, where we designate the systems with maximum stopping time $\tausmax = 0.2$ and $\tausmax = 2$ as Model~A and Model~B, respectively, and Model~A is further differentiated by its growth regime, f and s for fast and slow, respectively.
(We note that for our Model~B, the domain size is $2\Hg\times2\Hg$, which may have complications in the implications for dust-gas dynamics in the vertical direction; see Sections~\ref{SS:diff} and~\ref{SS:hp}.)

Following Paper~I, we use the \textsc{Pencil Code}\footnote{The \textsc{Pencil Code} and its documentation is publicly available at \url{http://pencil-code.nordita.org/}.} \citep{BD02,PCC21} to conduct the simulation models.
The linear growth of the instability for each selected case has been reproduced by the code.
Instead of constructing an exact eigenmode as the initial conditions, we use Gaussian noises in radial ($x$) and vertical ($z$) directions to perturb the positions of the particles on top of the equilibrium state such that the perturbation of the total particle density is $\delta\rhop \approx 10^{-6}\rho_{\mathrm{p},0}$, where $\rho_{\mathrm{p},0} \equiv \epsilon\rho_0$ is the mean density of the particles and $\rho_0$ is the mean density of the gas.
The gas density is initially uniform with $\rhog = \rho_0$.
The velocities of the gas $\ug$ and the $j$-th dust species $\vpj{j}$, with $j = 1, 2, ..., \Nsp$, are also uniform and in initial equilibrium $\ug = \vec{u}_{\mathrm{g}0}$ and $\vpj{j} = \vec{v}_{\mathrm{p}0,j}$, respectively, where $\Nsp$ is the number of discrete dust species in the dust-size distribution \citep{BS10c,BKP19}.
The code configuration is otherwise the same as in Paper~I.

We run the simulations for at least several $e$-folding times of the instability till the nonlinear saturation is obtained, and we systematically vary the number of discrete dust species $\Nsp$ with 1, 4, 16, and 64.
The maximum resolution we have been able to achieve is either $512\times512$ or $256\times256$ (Table~\ref{T:specs}), and in Appendix~\ref{S:res} we show the resolution study.
Unless otherwise noted, we report our resulting data at the highest resolution.

%-----------------------------------------------------------------------
\section{Saturated state and the gas} \label{S:sat}

We first construct several diagnostics to monitor the systems from linear growth to nonlinear saturation.
Following the numerical validation conducted in Sections~2.2 and~3.3 of Paper~I, we define the density fluctuation in the gas as a function of time $t$ by
\begin{equation} \label{E:drhog}
	\drhog \equiv \sqrt{\va{\rhog^2} - \va{\rhog}^2},
\end{equation}
where
\begin{equation}
    \va{f} \equiv \frac{1}{L_x L_z}\iint f\,\diff{x}\diff{z}
\end{equation}
denotes the volume average of $f$ over the computational domain at any given $t$.
In our models, $\va{\rhog} = \rho_0$ is maintained to at least seven significant digits.
We next define the mean velocity deviation of the gas as a function of $t$ by
\begin{equation} \label{E:Dug}
	\overline{\Dug} \equiv \frac{1}{\rho_0}\va{\rhog\Dug},
\end{equation}
where $\Dug \equiv \ug - \vec{u}_{\mathrm{g}0}$ is the velocity deviation of the gas from the initial equilibrium velocity $\vec{u}_{\mathrm{g}0}$.
Then the components of the velocity dispersion of the gas as a function of $t$ can be defined as
\begin{equation} \label{E:dug}
	\dugi{i} \equiv \sqrt{\frac{1}{\rho_0}\va{\rhog\Dugi{i}^2} - \ugmi{i}^2}.
\end{equation}
For the dust particles, we define the ensemble-averaged density fluctuation as a function of $t$ by
\begin{equation}
	\delta\rhop \equiv \sqrt{\va{\rhop^2} - \va{\rhop}^2}
\end{equation}
on the grid assigned by the particle-mesh method, where $\va{\rhop} = \epsilon\rho_0$ by construction.
With the Lagrangian super-particle approach, on the other hand, the components of the ensemble-averaged velocity dispersion of the dust particles are defined as
\begin{equation}
    \dvpi{i} \equiv
    \sqrt{\left(\frac{1}{M_\mathrm{p}}\sum_k\mpk{k}(\Dvpki{k}{i})^2\right) -
          \left(\frac{1}{M_\mathrm{p}}\sum_k\mpk{k}\Dvpki{k}{i}\right)^2},
\end{equation}%
\begin{landscape}
\begin{table}
\centering
\caption{Time-averaged properties of the gas at saturation state.
	The columns are
		(1)~model identifier,
		(2)~number of discrete dust species,
		(3)~estimated saturation time,
		(4)~density fluctuation,
		(5)--(6)~components of the velocity deviation from the initial equilibrium,
		(7)--(9)~components of the velocity dispersion, and
		(10)--(12)~components of the Reynolds stress, measured with respect to the mean velocity at saturation, $\vec{u}_\mathrm{g}' \equiv \Delta\vec{u}_\mathrm{g} - \overline{\Delta\vec{u}_\mathrm{g}}$.
	The time variability of the last digit is shown in parentheses.}
\label{T:turb}
\begin{tabular}{crcccccccccc}
	\hline
	Model & $\Nsp$ & $t_\mathrm{sat}$$^a$ & $\drhog$
	& $\ugmi{x}$     & $\ugmi{y}$
	& $\dugi{x}$     & $\dugi{y}$     & $\dugi{z}$
	& $\va{\rhog\ugi{x}'\ugi{y}'}$
	& $\va{\rhog\ugi{x}'\ugi{z}'}$
	& $\va{\rhog\ugi{y}'\ugi{z}'}$\smallskip\\
                &        & ($P$)            & ($\rho_0$)
	& ($\cs$) & ($\cs$) & ($\cs$) & ($\cs$) & ($\cs$)
	& ($\rho_0\cs^2$) & ($\rho_0\cs^2$) & ($\rho_0\cs^2$)\smallskip\\
	(1) & (2) & (3) & (4) & (5)	& (6) & (7) & (8)& (9) & (10) & (11) & (12)\\
	\hline
%
%.../taus31/eps2/nsp1/nx512
	Af    &  1
	& \phn6
	&       \sn{7.7(5)}{-5}
	&  \phn\sn{+2.5(1)}{-3} &       \sn{-5.0(6)}{-4}
	&   \phn\sn{7.3(2)}{-3} &    \phn\sn{2.7(2)}{-3} & \phn\sn{5.4(2)}{-3}
	&         \sn{-8.7}{-6} &          \sn{+8.0}{-8} & \sn{-3.2}{-8}\\
%.../taus31/eps2/nsp4/nx512
	            &  4
	&    12
	&       \sn{5.1(3)}{-5}
	&  \phn\sn{+9.5(5)}{-4} &       \sn{-1.6(3)}{-4}
	&   \phn\sn{5.0(2)}{-3} &    \phn\sn{2.2(2)}{-3} & \phn\sn{4.1(2)}{-3}
	&         \sn{-4.1}{-6} &          \sn{+3.4}{-8} & \sn{-1.8}{-8}\\
%.../taus31/eps2/nsp16/nx512
	            & 16
	&    24
	&       \sn{5.8(4)}{-5}
	&  \phn\sn{+8.2(5)}{-4} &       \sn{-2.1(3)}{-4}
	&   \phn\sn{5.1(2)}{-3} &    \phn\sn{2.5(2)}{-3} & \phn\sn{4.4(2)}{-3}
	&         \sn{-4.2}{-6} &          \sn{+9.6}{-8} & \sn{-4.3}{-8}\\
%.../taus31/eps2/nsp64/nx512
	            & 64
	&    24
	&       \sn{6.1(4)}{-5}
	&  \phn\sn{+8.2(6)}{-4} &       \sn{-2.3(4)}{-4}
	&   \phn\sn{5.2(2)}{-3} &    \phn\sn{2.6(2)}{-3} & \phn\sn{4.7(2)}{-3}
	&         \sn{-4.3}{-6} &          \sn{-5.7}{-8} & \sn{-4.1}{-8}\smallskip\\
%
%.../taus31/eps0.2/nsp1/nx256
	As    &  1
	& \phn100
	&       \sn{1.5(1)}{-4}
	&  \phn\sn{+5.7(3)}{-4} &       \sn{-2.2(3)}{-4}
	&      \sn{1.06(3)}{-2} &    \phn\sn{6.4(2)}{-3} & \sn{1.00(3)}{-2}
	&         \sn{-1.4}{-5} &          \sn{-5.6}{-8} & \sn{+1.1}{-7}\\
%.../taus31/eps0.2/nsp4/nx256
	            &  4
	& \phn600
	&       \sn{3.2(2)}{-6}
	&  \phn\sn{+4.7(1)}{-6} &       \sn{-2.7(1)}{-6}
	&   \phn\sn{7.7(1)}{-4} &       \sn{5.70(8)}{-4} & \sn{2.34(9)}{-3}
	&         \sn{-7.4}{-8} &         \sn{+2.8}{-10} & \sn{+1.5}{-9}\\
%.../taus31/eps0.2/nsp16/nx256
	            & 16
	& 2800
	& \phd\phn\sn{3(2)}{-6}
	&     \sn{+1.11(8)}{-6} & \phd\phn\sn{-9(3)}{-7}
	&   \phn\sn{4.6(4)}{-4} &    \phn\sn{3.4(7)}{-4} & \phn\sn{2.2(3)}{-3}
	&         \sn{-1.6}{-8} &         \sn{+9.1}{-10} & \sn{-9.6}{-10}\\
%.../taus31/eps0.2/nsp64/nx256
	            & 64
	& 3000
	& \phd\phn\sn{4(1)}{-6}
	&  \phn\sn{+2.3(4)}{-7} &       \sn{-3.1(6)}{-7}
	&   \phn\sn{2.7(2)}{-4} &    \phn\sn{2.3(3)}{-4} & \phn\sn{3.0(6)}{-4}
	&         \sn{-3.3}{-9} &          \sn{-4.8}{-9} &      \sn{-1.3}{-10}\smallskip\\
%
%.../taus30/eps0.2/nsp1/nx512
	B     &  1
	& \phn60
	&       \sn{3.4(6)}{-3}
	&  \phn\sn{-3.0(6)}{-3} &       \sn{-1.5(3)}{-3}
	&   \phn\sn{1.2(2)}{-2} &    \phn\sn{1.8(1)}{-2} & \phn\sn{3.9(3)}{-2}
	&         \sn{+2.4}{-5} &          \sn{+1.8}{-5} & \sn{+7.4}{-6}\\
%.../taus30/eps0.2/nsp4/nx512
	            &  4
	& \phn70
	&       \sn{2.6(2)}{-3}
	&  \phn\sn{-1.6(2)}{-3} &       \sn{-4.9(9)}{-4}
	&      \sn{1.11(5)}{-2} &       \sn{1.46(7)}{-2} & \phn\sn{3.2(2)}{-2}
	&         \sn{+1.5}{-5} &          \sn{-1.5}{-5} & \sn{-1.9}{-6}\\
%.../taus30/eps0.2/nsp16/nx512
	            & 16
	& 100
	&       \sn{1.9(2)}{-3}
	&  \phn\sn{-1.1(2)}{-3} &       \sn{-4.0(7)}{-4}
	&   \phn\sn{8.5(4)}{-3} &       \sn{1.13(9)}{-2} & \phn\sn{2.3(2)}{-2}
	&         \sn{+1.4}{-5} &          \sn{+9.9}{-6} & \sn{+3.4}{-6}\\
%.../taus30/eps0.2/nsp64/nx512
                & 64
	& 140
	&       \sn{2.1(2)}{-3}
	&     \sn{-1.20(9)}{-3} &       \sn{-4.4(6)}{-4}
	&   \phn\sn{8.4(3)}{-3} &       \sn{1.22(5)}{-2} & \phn\sn{2.6(1)}{-2}
	&         \sn{+1.4}{-5} &          \sn{-2.3}{-5} & \sn{-5.9}{-6}\\
	\hline
	\multicolumn{12}{l}{$^a$~All time averages reported in this work (including both for the gas and for the particles) are conducted from $t_\mathrm{sat}$ to $t_\mathrm{max}$ of Table~\ref{T:specs}.}
\end{tabular}
\end{table}
\end{landscape}
\noindent
where $M_\mathrm{p} \equiv \sum_k\mpk{k}$ is the total mass of the dust particles, $\mpk{k}$ is the mass of the $k$-th individual dust super-particle, $\Delta\vpk{k} \equiv \vpk{k} - \vec{v}_{\mathrm{p}0,\texttt{spec}(k)}$ is the velocity deviation of the $k$-th super-particle from the initial equilibrium velocity, and $\texttt{spec}(k)$ is the dust species the $k$-th super-particle belongs to.
With these definitions, the evolution of the density fluctuations and the velocity dispersions of the gas and the dust particles for each dust-size distribution listed in Table~\ref{T:specs} is shown in Figs.~\ref{F:disp1}--\ref{F:disp3}, respectively.

\begin{figure*}
	\includegraphics[width=0.9\textwidth]{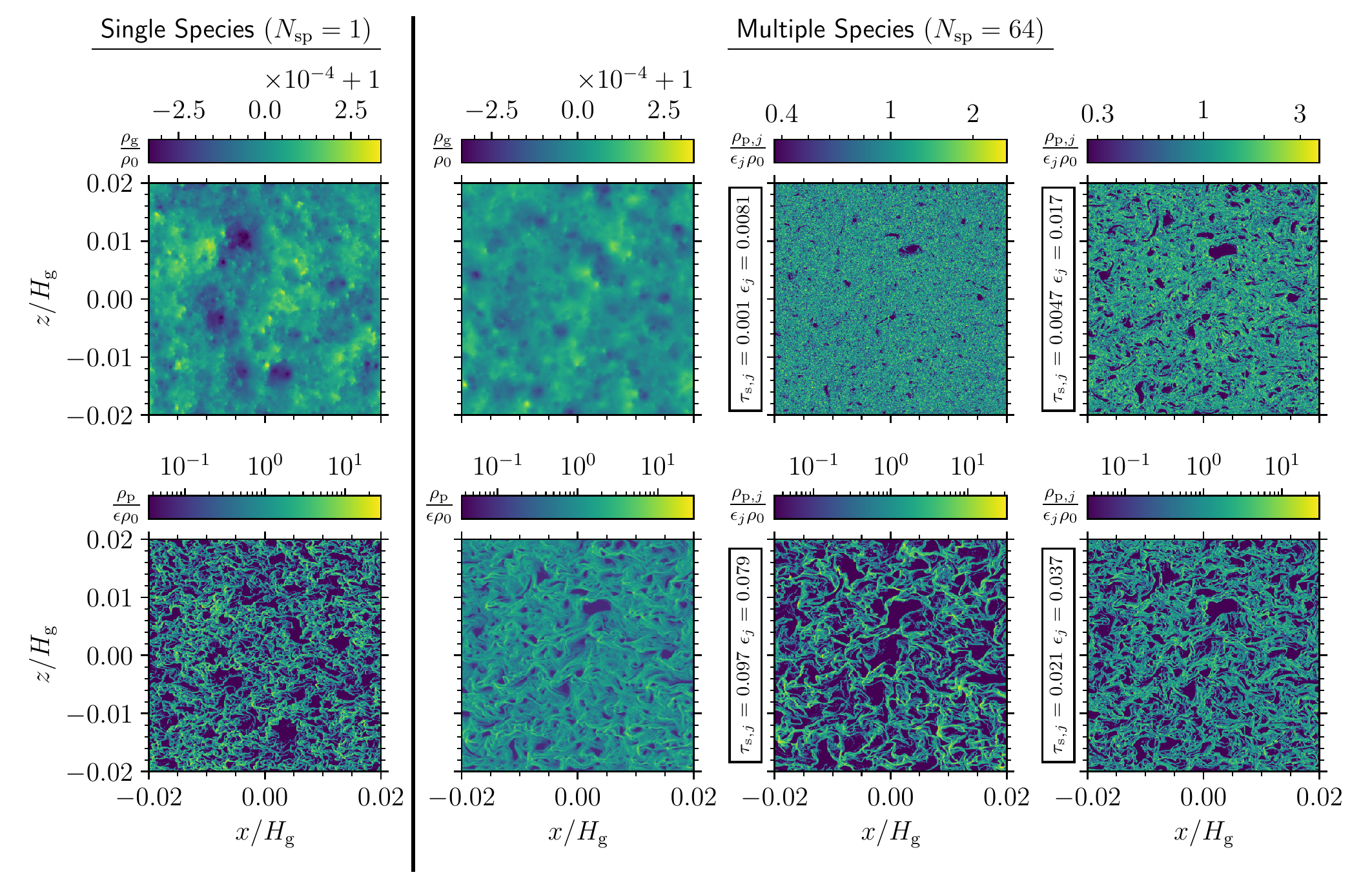}
	\caption{Densities at the end of the simulations for Model~Af (\dda{} with $\epsilon = 2$).
	    The first column is the case of single species, while the other three columns are the case of multiple species with the number of dust species $\Nsp = 64$.
		On the left two columns, the top and the bottom panels show the gas density and the total density of dust particles, respectively.
		On the right two columns, the densities of four individual dust species are presented, clockwise with increasing $\taus$.
		All the densities are normalised by their initial mean densities.}
	\label{F:dens1}
\end{figure*}

\begin{figure*}
	\includegraphics[width=0.9\textwidth]{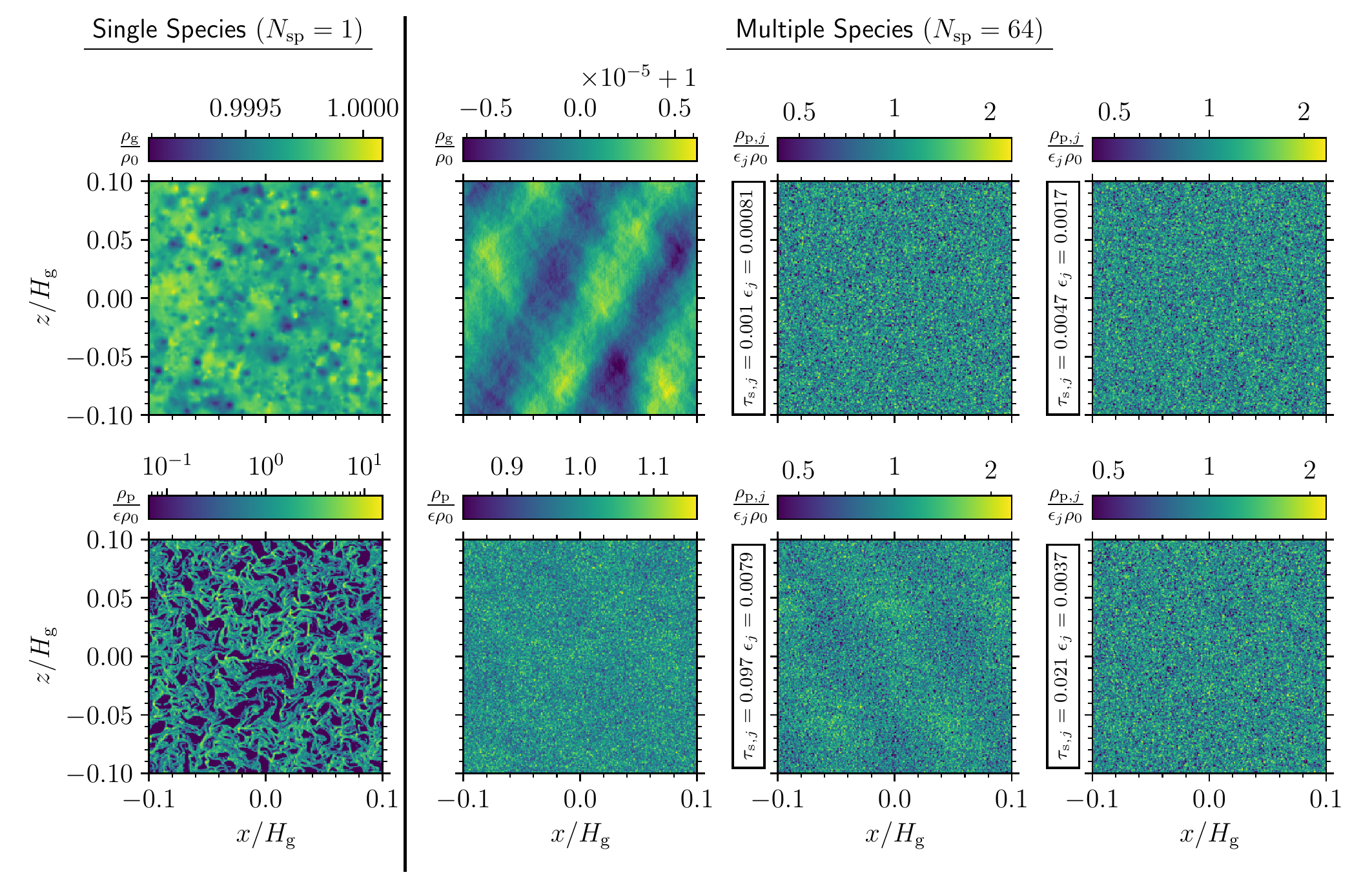}
	\caption{Similar to Fig.~\ref{F:dens1} except for Model~As (\dda{} with $\epsilon = 0.2$).}
	\label{F:dens2}
\end{figure*}

\begin{figure*}
	\includegraphics[width=0.9\textwidth]{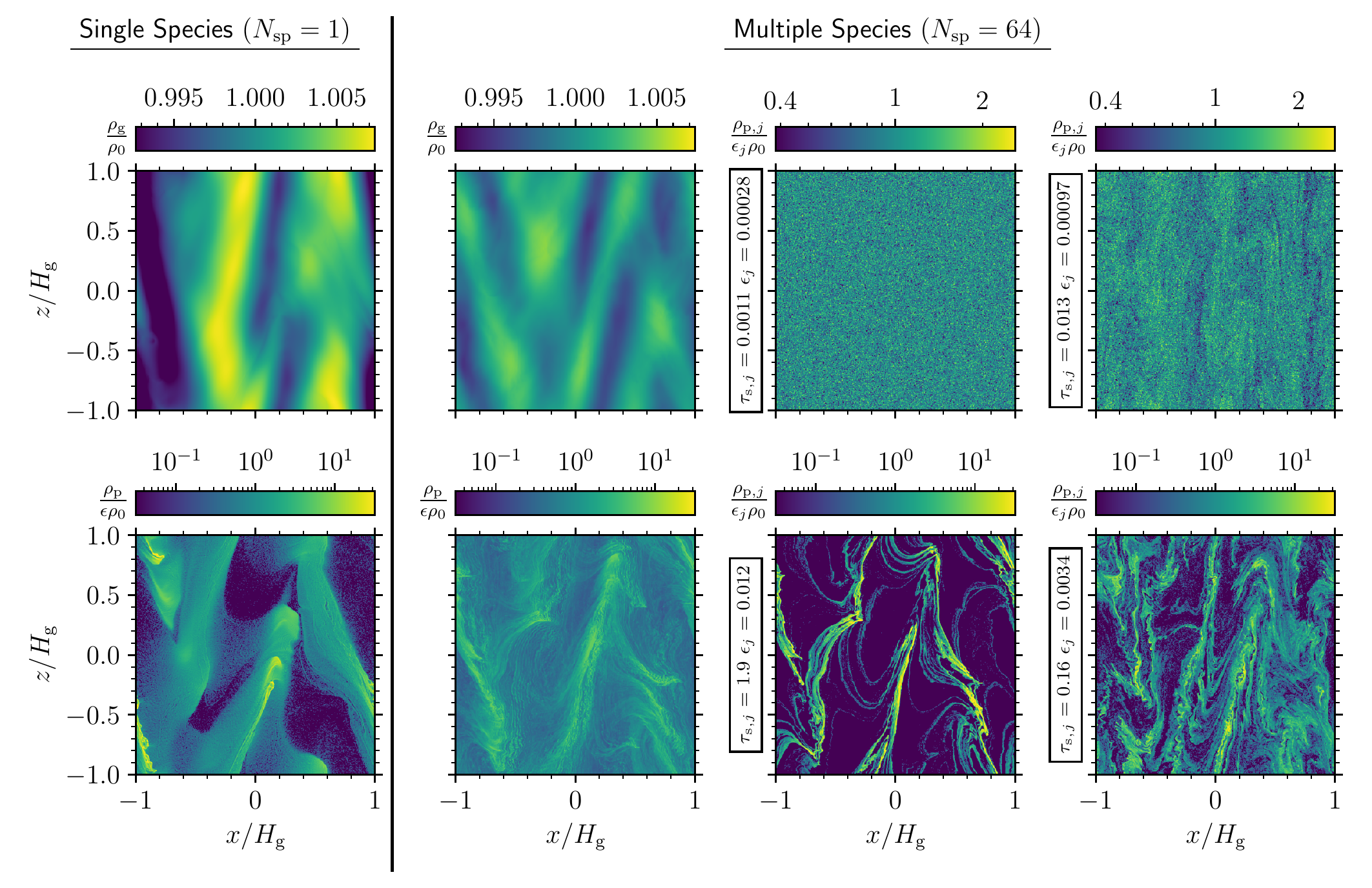}
	\caption{Similar to Fig.~\ref{F:dens1} except for Model~B (\ddb{} with $\epsilon = 0.2$).}
	\label{F:dens3}
\end{figure*}

In all cases, the evolution of the diagnostics begins with an initial exponential growth, subsequently undergoes a transition period, and finally levels off and reaches a nonlinear saturation state.
We estimate the points of saturation by inspecting the figures and record the times in Table~\ref{T:turb}.
It is apparent that the saturation time correlates with the growth rate of the linear phase of the streaming instability.
For the two distributions categorised as in the fast-growth regime, the growth rate generally decreases with the number of discrete dust species $\Nsp$ in the distribution until $\Nsp \sim 16$--64, at which the growth rate obtains convergence (Paper~I).
We note that the growth rates of Model~Af with $\Nsp \gtrsim 16$ and Model~B with $\Nsp \gtrsim 64$ approximate those in the continuum limit ($\Nsp \rightarrow \infty$) to within 3\% (S.-J.~Paardekooper, private communication).
For Model~Af shown in Fig.~\ref{F:disp1}, the case of $\Nsp = 16$ has essentially the same evolution as the case of $\Nsp = 64$, while for Model~B shown in Fig.~\ref{F:disp3}, the case of $\Nsp = 16$ closely approaches the case of $\Nsp = 64$.
On the other hand, for Model~As shown in Fig.~\ref{F:disp2}, which is in the slow-growth regime, the growth rate continues to decrease significantly and hence the saturation time continues to increase, up to our highest $\Nsp = 64$.
As discussed in Paper~I, it remains unclear for this case how many dust species are required to approach the continuum limit and if the growth rate is finite at all (no positive rate has been found in the continuum limit for this case using the method of \citealt{PML21}; S.-J.~Paardekooper, private communication), but it is evident that the rate, if any, is problematically low as compared to the orbital timescale.
It is still computationally infeasible to simulate this kind of systems with $\Nsp \gtrsim O(10^2)$, and hence we base our discussion on the trend seen with $\Nsp \lesssim 64$.
The gas and dust densities at the end of each simulation with $\Nsp = 64$ are shown in Figs.~\ref{F:dens1}--\ref{F:dens3}, respectively.

With the estimated saturation times $t_\mathrm{sat}$, we focus on the gas and report in Table~\ref{T:turb} the time averages of several properties at the saturation state.
We report in this work all time averages conducted from $t_\mathrm{sat}$ to the end of the simulations $t_\mathrm{max}$ (Table~\ref{T:specs}).
First of all, the fluctuation in gas density $\drhog$ (equation~\eqref{E:drhog}) is in general small ($10^{-6} \lesssim \drhog / \rho_0 \lesssim 10^{-3}$), indicating a high degree of incompressibility in the gas component, consistent with the nonlinear saturation of single-species streaming instability \citep{JY07}.
For the case with the largest dust species $\tausmax = 2$ (Model~B), $\drhog \sim 10^{-3}\rho_0$ is considerably larger than that for the case with $\tausmax = 0.1$ (Models~Af and~As).
Comparing the top and bottom panels on the left two columns of Figs.~\ref{F:dens1}--\ref{F:dens3}, we note that the gas density fluctuation bears similar pattern as the dust density fluctuation, indicating the importance of the dust back reaction to the gas.

For Model~As, which is in the slow-growth regime, we note that the saturation of the multi-species streaming instability behaves appreciably differently from that of the single-species.
The latter reaches a much stronger disturbed state with $\drhog \sim 10^{-4}\rho_0$, similar in dynamics as its high-$\epsilon$ counterpart as shown in Fig.~\ref{F:dens1} \citep[as well as Models~AB and~AC of][]{JY07}.
On the other hand, when $\Nsp > 1$, the system reaches a significantly quiescent state with $\drhog \sim 10^{-6}\rho_0$, as shown in Fig.~\ref{F:dens2}.
The same dichotomy is also seen in all other measured properties listed in Table~\ref{T:turb}, and hence we exclude the single-species version of this distribution from the discussion below in this section.
We present in Section~\ref{S:dust} this dichotomy in further detail.
It is noticeable in Fig.~\ref{F:disp2} that there exist oscillations in our diagnostics at the saturation state when $\Nsp = 64$.
Given that the level of these dispersions is so low, these oscillations could be numerical noise (see also Section~\ref{SS:diff}), however they do not alter our measurements using time averages.

The fifth and the sixth columns in Table~\ref{T:turb} lists the horizontal components of the mean gas velocity deviation $\overline{\Dug}$ from the initial equilibrium velocity (equation~\eqref{E:Dug}).
It appears that the turbulent saturation state does shift the gas velocity from the initial equilibrium.
On one hand, the gas becomes slightly \emph{more} sub-Keplerian in general.
For the fast-growth regime, the deviation is on the order of $10^{-4}\cs$, where $\cs$ is the speed of sound, as compared to the dust-free sub-Keplerian speed at $\Pi\cs = 0.05\cs$ (Section~\ref{S:method}).
For the slow-growth regime, the deviation approaches zero and no convergence to a finite value is seen with increasing number of dust species up to $\Nsp = 64$.
On the other hand, the radial speed of the gas is either enhanced or reduced from the initial \emph{outward} radial motion, depending on the largest dust species.
For $\tausmax = 0.1$ (Models~Af and~As), the radial motion of the gas speeds up, the magnitude of which depends on the regime.
For $\tausmax = 2$ (Model~B), the radial motion of the gas slows down.

In addition to the mean velocity deviation from the initial equilibrium, the gas also demonstrates appreciable velocity dispersion $\delta\vec{u}_\mathrm{g}$ (equation~\eqref{E:dug}), as shown in Columns~(7)--(9) of Table~\ref{T:turb}.
For Model~Af, all components are on the order of $10^{-3}\cs$ with $\dugi{y} \lesssim \dugi{z} \lesssim \dugi{x}$.
We note that $\dugi{y} \sim \dugi{x} / 2$ in this case, consistent with epicycle motions executed by the gas \citep[see, e.g.,][]{PT06,YMM09}.
For Model~B, all components are on the order of $10^{-2}\cs$ with $\dugi{x} \lesssim \dugi{y} \lesssim \dugi{z}$.
On the other hand, for Model~As, all components decreases with increasing $\Nsp$ and no convergence to a finite value is seen.

Based on all the properties measured above, we compute the components of the Reynolds stress.
The fluctuating part of the gas velocity $\vec{u}_\mathrm{g}'$ has been evaluated against the mean velocity at the saturated state \citep{aB98}: $\vec{u}_\mathrm{g}' \equiv \Delta\vec{u}_\mathrm{g} - \overline{\Delta\vec{u}_\mathrm{g}}$.
The results are listed in the last three columns of Table~\ref{T:turb}.

The horizontal component of the stress $\va{\rhog\ugi{x}'\ugi{y}'}$, which indicates the transport of angular momentum, is particularly interesting.
For the fast-growth regime, its sign depends on the largest dust species.
With $\tausmax = 0.1$ (Model~Af), the stress is \emph{negative}, leading to regular turbulent viscosity with magnitude on the order of $10^{-6}\cs\Hg$.
By contrast, with $\tausmax = 2$ (Model~B), the stress is \emph{positive}, indicating that the gas undergoes anti-diffusion.
From the left two columns of Fig.~\ref{F:dens3}, it can be seen that the gas tends to concentrate in \emph{between} the dust filaments.
This phenomenon has also been seen in the simulations of single-species streaming instability with vertical gravity, when strong radial concentration of dust particles occurs, and was discussed by \cite{YJ14}.
We note that however, the anti-correlation observed in vertically stratified systems may depend on the vertical boundary conditions and can be significantly weakened with increasing vertical dimension \citep{YJ14,LYS18}.
As for the slow-growth regime (Model~As), the stress is negative and significantly smaller, and no convergence to a finite value is seen with increasing $\Nsp$.

Finally, it appears that there is no general trend in the other two components of the Reynolds stress, $\va{\rhog\ugi{x}'\ugi{z}'}$ and $\va{\rhog\ugi{y}'\ugi{z}'}$, either in sign or in magnitude.
It may result from the indeterminism of the vertical velocity, since there exists no restoring force in the vertical direction.
We note that for Model~B, there is significant vertical motion, leading to appreciable $\va{\rhog\ugi{x}'\ugi{z}'}$ and $\va{\rhog\ugi{y}'\ugi{z}'}$ stresses.
In any case, these components should be further quantified by introducing vertical gravity in the system.

%-----------------------------------------------------------------------
\section{Dust Dynamics} \label{S:dust}

In this section, we focus on the dust dynamics at the saturation state of multi-species streaming instability.
Specifically, density distribution, kinematics, and turbulent diffusion of each dust species are analysed.
We also discuss some implications of the results for vertical scale height and radial transport of the dust particles.

%-----------------------------------------------------------------------
\subsection{Density distribution} \label{SS:dd}

The first column in Figs.~\ref{F:dens1}--\ref{F:dens3} shows the total densities of gas and dust at the end of the simulations with single species $\Nsp = 1$.
The saturation state with single species recovers the two distinct dynamics found in the literature \citep{JY07,BS10a,YJ16,BKP19}.
When the dimensionless stopping time $\taus \lesssim 0.1$ as in Figs.~\ref{F:dens1} and~\ref{F:dens2}, the system evolves into a turbulent state of numerous vortices with dust particles trapped in between, similar to Model~AB in \cite{JY07}.
This occurs irrespective of the solid-to-gas density ratio $\epsilon$.
When $\taus \gtrsim 1$ as in Fig.~\ref{F:dens3}, by contrast, the dust collects themselves in the radial direction analogous to traffic jams into axisymmetric filaments while moving freely in the vertical direction, similar to Model~BA in \cite{JY07}.
In all three cases, the relative fluctuation in total dust density is on the order of unity (Figs.~\ref{F:disp1}--\ref{F:disp3}).

With multiple species, however, the saturation state can differ depending on the total solid-to-gas density ratio $\epsilon$.
The second column in Figs.~\ref{F:dens1}--\ref{F:dens3} shows the total densities of gas and dust at the end of the simulations with multiple species $\Nsp = 64$.
When in the fast-growth regime of the streaming instability (Models~Af and~B), the system appears to be similar to its single-species counterpart, as shown in Figs.~\ref{F:dens1} and~\ref{F:dens3}, including the relative fluctuation in total dust density (Figs.~\ref{F:disp1} and~\ref{F:disp3}).
By contrast, when in the slow-growth regime (Model~As), the saturation state is significantly less turbulent and appears to be close to laminar, with a relative fluctuation of $\sim$10$^{-6}$ in gas density (see also Table~\ref{T:turb}) and $\simeq$4\% in total dust density, as shown in Figs.~\ref{F:disp2} and~\ref{F:dens2}.

\begin{figure*}
	\centering
	\subcaptionbox{Model~Af (\dda{} with $\epsilon = 2$)\label{F:vort1}}
		{\includegraphics[width=\textwidth]{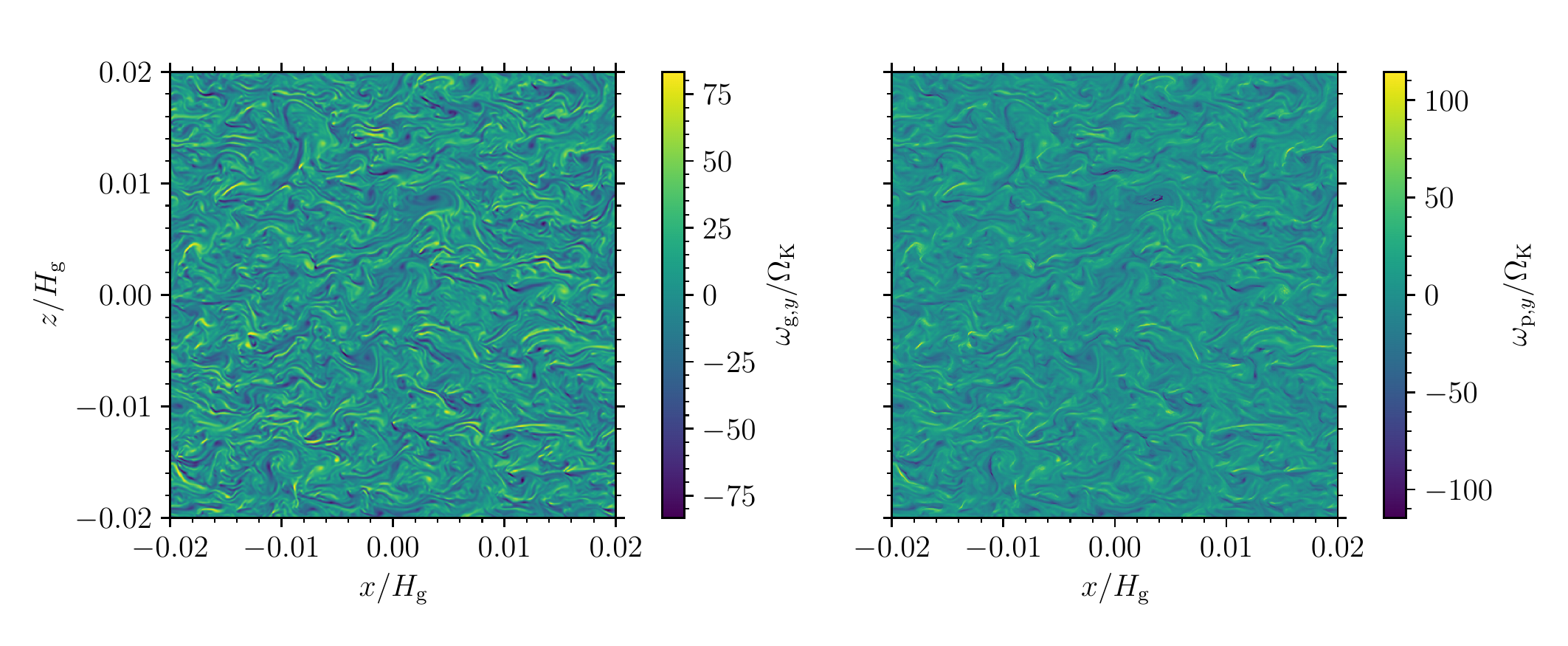}}
	\subcaptionbox{Model~B (\ddb{} with $\epsilon = 0.2$)\label{F:vort3}}
		{\includegraphics[width=\textwidth]{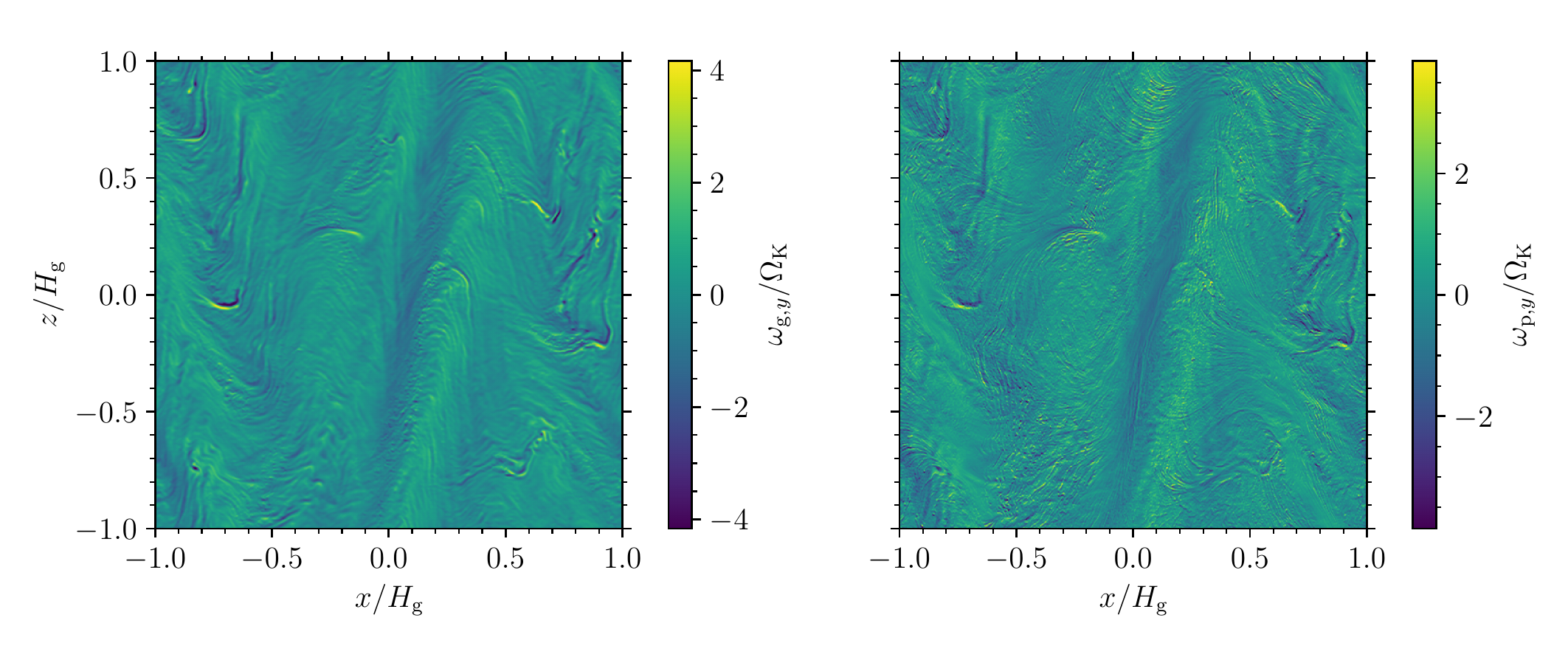}}
	\caption{Azimuthal component of the vorticity of the gas (\textit{left panel}) and the dust particles (\textit{right panel}) at the end of two models with $\Nsp = 64$ dust species.
	    The vorticity is normalised by the Keplerian angular frequency $\Omega_\mathrm{K}$.}
	\label{F:vort}
\end{figure*}

To demonstrate the prevalence of vortices in Model~Af discussed above, we plot in Fig.~\ref{F:vort} the azimuthal component of the vorticity of the gas and the dust particles at the end of Models~Af and~B with $\Nsp = 64$.
The vorticity of the gas is $\vec{\omega}_\mathrm{g} \equiv \vec{\nabla}\times\ug$, while we compute the vorticity of the dust particles using their total velocity via $\vec{\omega}_\mathrm{p} \equiv \vec{\nabla} \times \vec{V}_\mathrm{p}$, where $\vec{V}_\mathrm{p} \equiv \sum_k (\rhop \vec{v}_\mathrm{p})^{(k)} / \sum_k \rhop^{(k)}$ and the superscript $(k)$ denotes the contribution by the $k$-th super-particle to a cell assigned by the particle-mesh method.
As shown in Fig.~\ref{F:vort1}, numerous small-scale but strong vortices can be identified in Model~Af, for instance, at $(x, z) = (+0.0039\Hg, +0.0085\Hg)$ and $(-0.0063\Hg, +0.0134\Hg)$ in opposite circulation, where they coincide with the depression of the gas and dust densities in Fig.~\ref{F:dens1}.
On the contrary, the vorticity in Model~B is significantly weaker, as shown in Fig.~\ref{F:vort3}, and no conspicuous vortex structure can be well defined.

We further decompose the dust density by individual species, as shown in the right two columns of Figs.~\ref{F:dens1}--\ref{F:dens3}.
The panels are arranged such that the stopping time $\taus$ increases in the clockwise direction.
For the slow-growth regime shown in Fig.~\ref{F:dens2}, this decomposition appears less interesting, with small fluctuations in dust density for all species.
For the fast-growth regime, on the other hand, we find that dust particles of different $\taus$ do demonstrate distinguishably different dynamics.
When the system undergoes vortical motions as in Fig.~\ref{F:dens1}, smaller particles tend to be more diffuse, while larger particles tend to be more collected in between vortices.
This difference is even more pronounced when the system undergoes radial traffic jams as in Fig.~\ref{F:dens3}.
The distribution of the smallest particles are virtually uniform, while the largest particles are concentrated into dense axisymmetric dust filaments.
Therefore, it can be concluded that the large particles in a dust distribution execute the dynamics similar to its single-species counterpart (when in the fast-growth regime), while the small particles are more diffuse and uniform because they are more tightly coupled to the gas, which has a high degree of incompressibility.

\begin{figure}
	\centering
	\subcaptionbox{Model~Af\label{F:rpdf1}}
		{\includegraphics[width=\columnwidth]{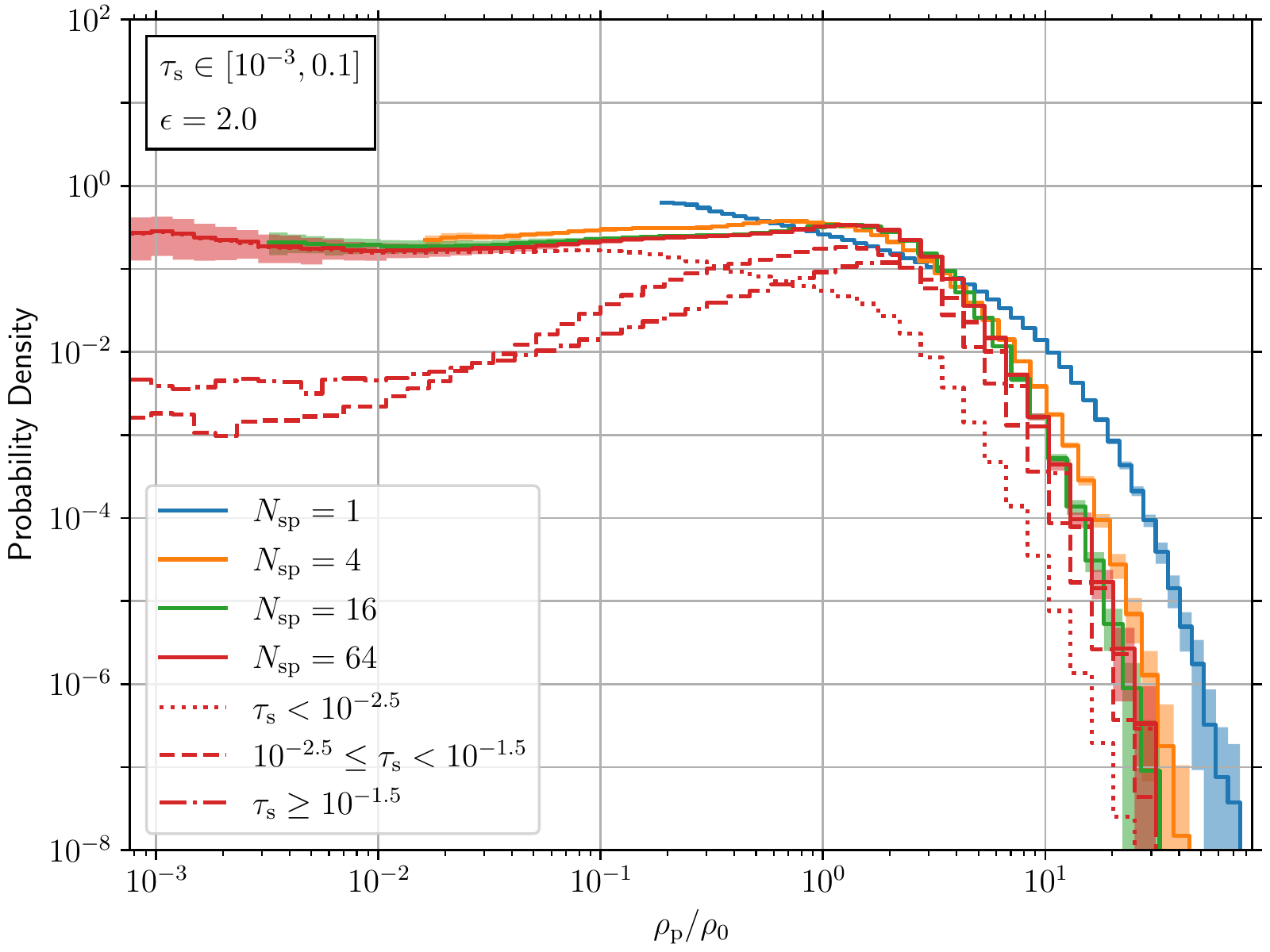}}
	\subcaptionbox{Model~As\label{F:rpdf2}}
		{\includegraphics[width=\columnwidth]{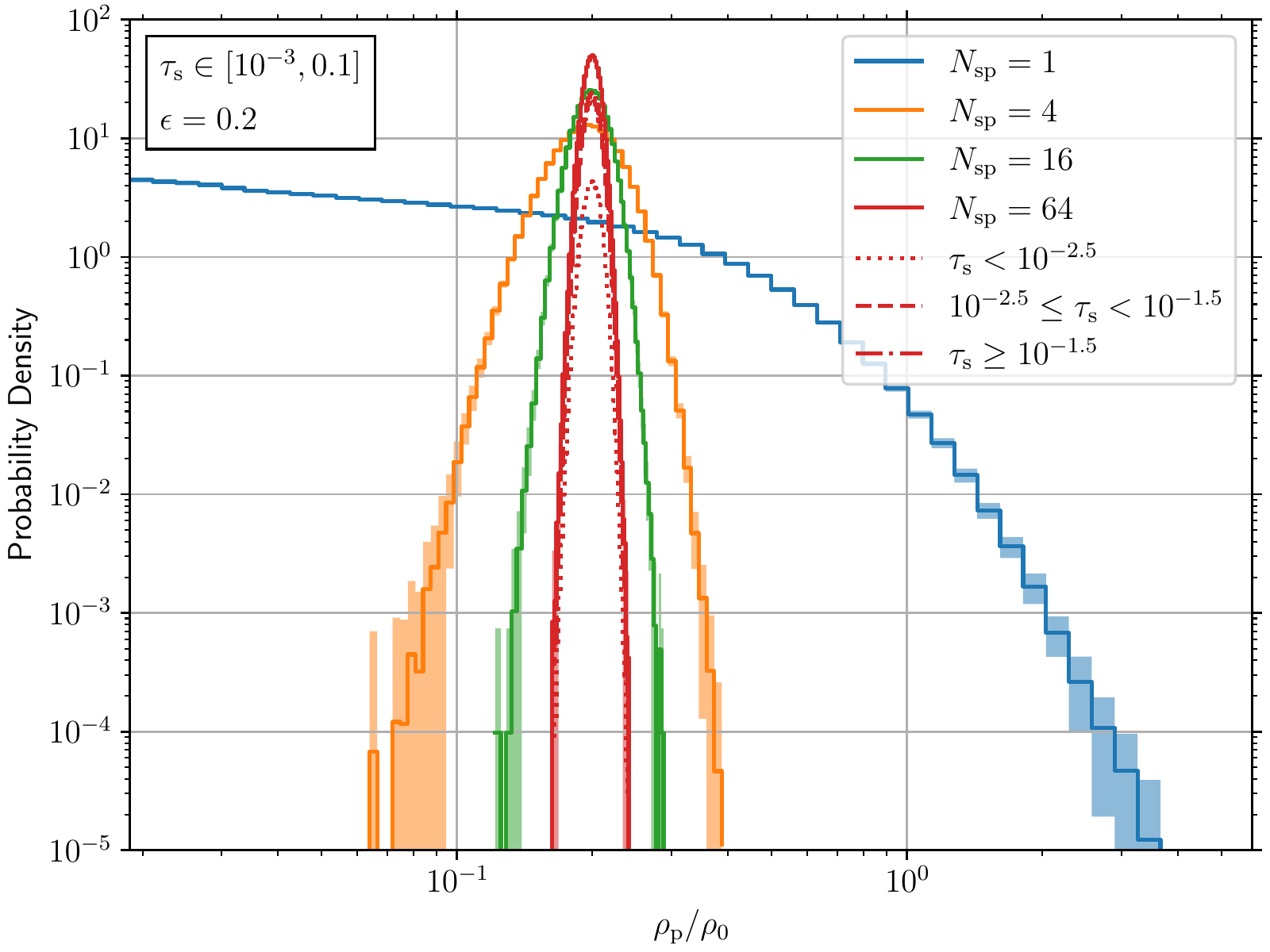}}
	\subcaptionbox{Model~B\label{F:rpdf3}}
		{\includegraphics[width=\columnwidth]{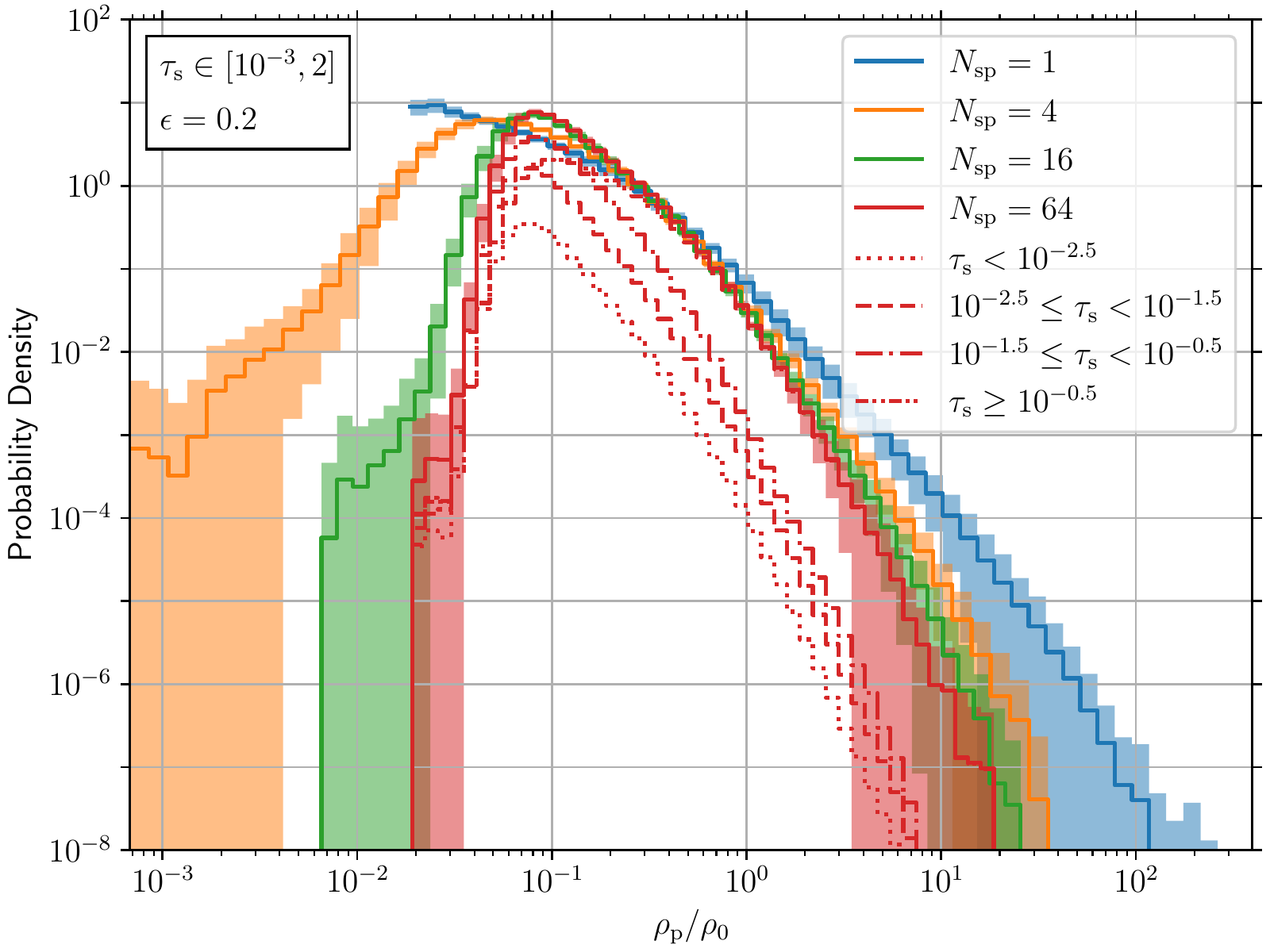}}
	\caption{Time-averaged probability density as a function of the total dust density $\rhop$.
	    Each solid line represents the model with a different number of dust species $\Nsp$, and the corresponding shade indicates the time variability.
	    The non-solid lines show the various contributions to the probability density by bins of dimensionless stopping time $\taus$ in decades for the model with $\Nsp = 64$.}
	\label{F:rpdf}
\end{figure}

With the spatial particle distribution as a function of time, we find the time-averaged probability density function (PDF) of total dust density $\rhop$ and plot it in Fig.~\ref{F:rpdf} for each model at the saturation state.
For Model~Af shown in Fig.~\ref{F:rpdf1}, $\rhop$ has a steep cutoff on the order of 10$\rho_0$ on the high end, where $\rho_0$ is the mean gas density, while levelling off toward lower density regions with $\rhop \lesssim 2\rho_0$, i.e., near and below the mean $\epsilon = 2$.
The PDF with single species ($\Nsp = 1$) has a higher density cutoff without sufficient sampling of low density regions, which is noticeably different from that with multiple species.
On the other hand, the PDFs between $\Nsp = 16$ and $\Nsp = 64$ are virtually the same except that the case of $\Nsp = 64$ can probe even lower density regions, indicating that convergence can be achieved with $\Nsp \gtrsim 16$.

For Model~As shown in Fig.~\ref{F:rpdf2}, which is in the slow-growth regime, it is evident that the PDF of total dust density $\rhop$ with single species is intrinsically different from the ones with multiple species.
As shown in Fig.~\ref{F:dens2}, the system with multiple species is close to quiescent while the one with single species remains turbulent as does its high-$\epsilon$ counterpart.
Therefore, the PDFs with $\Nsp > 1$ resembles a Gaussian distribution centred at the mean $\epsilon = 0.2$.
The width of the distribution becomes smaller with increasing $\Nsp$, and it is not clear if a finite width could be obtained with higher $\Nsp > 64$.

As noted above, Model~B undergoes different dust dynamics and hence results in distinctly different PDFs as shown in Fig.~\ref{F:rpdf3} as compared to Fig.~\ref{F:rpdf1}.
The PDFs in this case have an extended tail in high density regions while a steep cutoff on the low end.
As the number of dust species $\Nsp$ increases, the sharper the cutoff becomes, approaching $\rhop \sim 0.04\rho_0$ with $\Nsp = 64$, which in fact establishes a relatively uniform and diffuse background of small dust as exemplified in Fig.~\ref{F:dens3}.
Excluding the extreme high density tail where the Poisson noise dominates, the PDFs with $\Nsp = 16$ and $\Nsp = 64$ are consistent with each other near and above the peak at $\rhop \simeq 0.08\rho_0$, indicating numerical convergence.
We note that the time-averaged maximum dust density in these multi-species systems is $\max(\rhop) \sim 20\rho_0$, almost one order of magnitude smaller than that in their single-species counterpart.

%-----------------------------------------------------------------------
\subsection{Dust segregation and size distribution} \label{SS:sfd}

In Fig.~\ref{F:rpdf}, we further decompose the contribution of dust particles of different sizes to any given total dust density $\rhop$ with dotted and dashed red curves.
We bin the particles by decades of dimensionless stopping time $\taus$ and compute at each $\rhop$ bin the part of the total probability density attributed to the particles in each $\taus$ bin.
This is achieved by defining the mass-weighted probability of finding particles with $[\taus, \taus + \diff{\taus})$ in cells with total particle density $[\rhop, \rhop + \diff{\rhop})$ as
\begin{equation} \label{E:prob}
    p(\rhop,\taus)\diff{\rhop} \diff{\taus} \equiv
    \frac{1}{\rhop N_\mathrm{cell} V_\mathrm{cell}}
    \sum_{\substack{%
            \mathrm{cells} \in [\rhop, \rhop + \diff{\rhop}) \\
            \taus^{(k)} \in [\taus, \taus + \diff{\taus})}} \mpk{k},
\end{equation}
where $N_\mathrm{cell}$ is the total number of cells, $V_\mathrm{cell}$ is the volume of each cell, and $\taus^{(k)}$ is the dimensionless stopping time of the $k$-th super-particle.
In equation~\eqref{E:prob}, we use the weight assigned by the particle-mesh method to compute the mass of the $k$-th particle contributed to each cell $\mpk{k}$ for each $k$ to utilise the high spatial order of the triangular-shape-cloud scheme.
Only the results for models with number of species $\Nsp = 64$ are shown.

As demonstrated by Fig.~\ref{F:rpdf1}, dust particles of different sizes are segregated by the total dust density in Model~Af.
The large particles tend to be present in dense regions, while the small particles tend to be in diffuse regions (see also Fig.~\ref{F:dens1}).
On one hand, large particles with $\taus \gtrsim 10^{-1.5}$ dominate the PDF when $\rhop \gtrsim 3\rho_0$.
On the other hand, small particles with $\taus \lesssim 10^{-2.5}$ dominate the PDF when $\rhop \lesssim 0.3\rho_0$.
It appears that the vortices in this system slingshot the larger particles more easily and collect them in between, enhancing the dust density there \citep{CH01,PP11,HCW17,HC20}.
By contrast, no noticeable segregation is seen for the same dust distribution but with lower $\epsilon = 0.2$ (Model~As), which has a quiescent saturation state, as shown by Fig.~\ref{F:rpdf2}.

Model~B also exhibits some degree of dust segregation, as demonstrated by Fig.~\ref{F:rpdf3}.
The dense filamentary structures in its saturation state tend to be resided by the largest particles (Fig.~\ref{F:dens3}), and hence particles with $\taus \gtrsim 10^{-0.5}$ dominate the PDF when $\rhop \gtrsim 0.2\rho_0$.
The segregation is not as prominent as the other turbulent system such as Model~Af, though.

\begin{figure}
	\centering
	\subcaptionbox{Model~Af\label{F:sfd1}}
		{\includegraphics[width=\columnwidth]{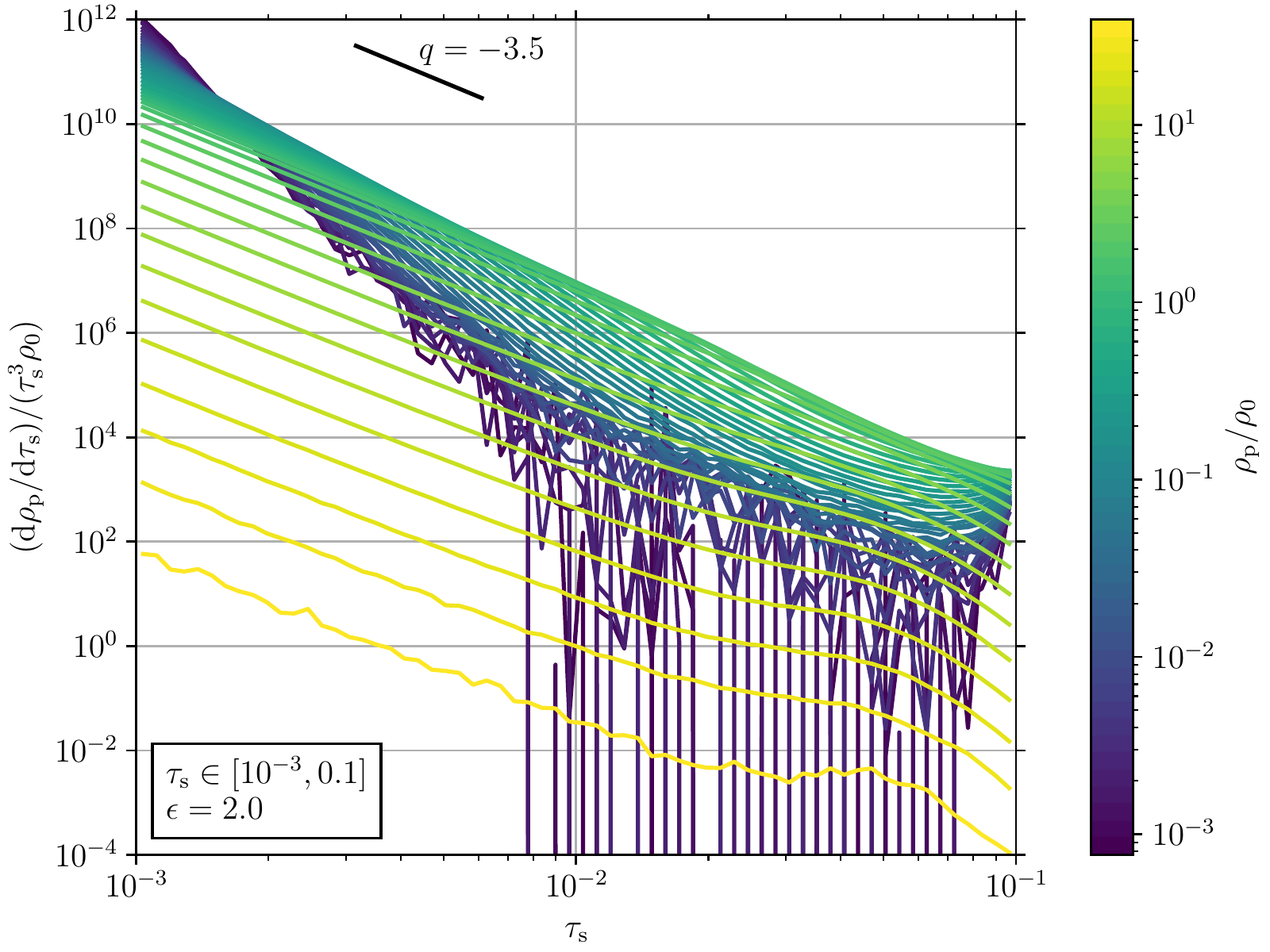}}
	\subcaptionbox{Model~As\label{F:sfd2}}
		{\includegraphics[width=\columnwidth]{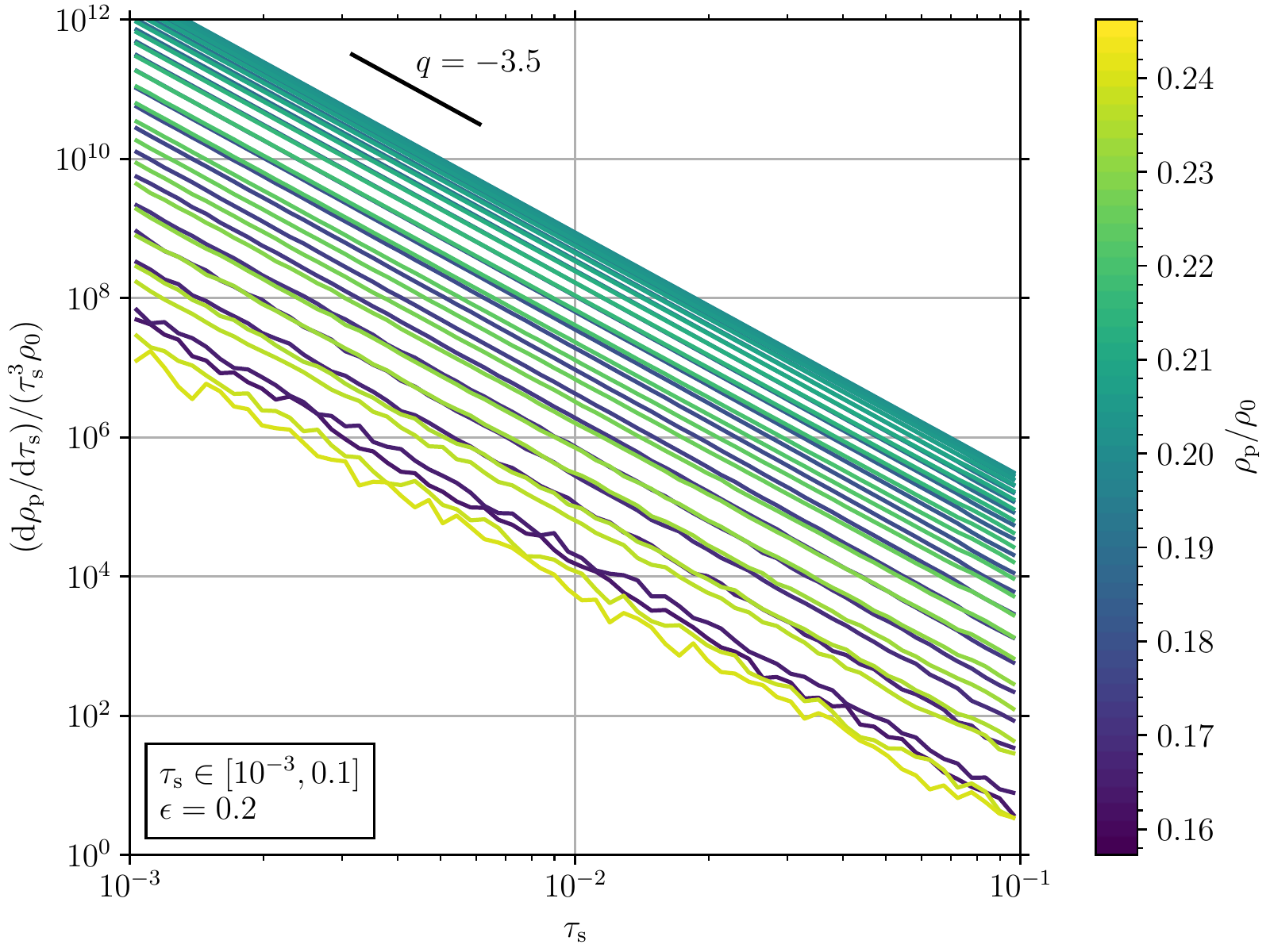}}
	\subcaptionbox{Model~B\label{F:sfd3}}
		{\includegraphics[width=\columnwidth]{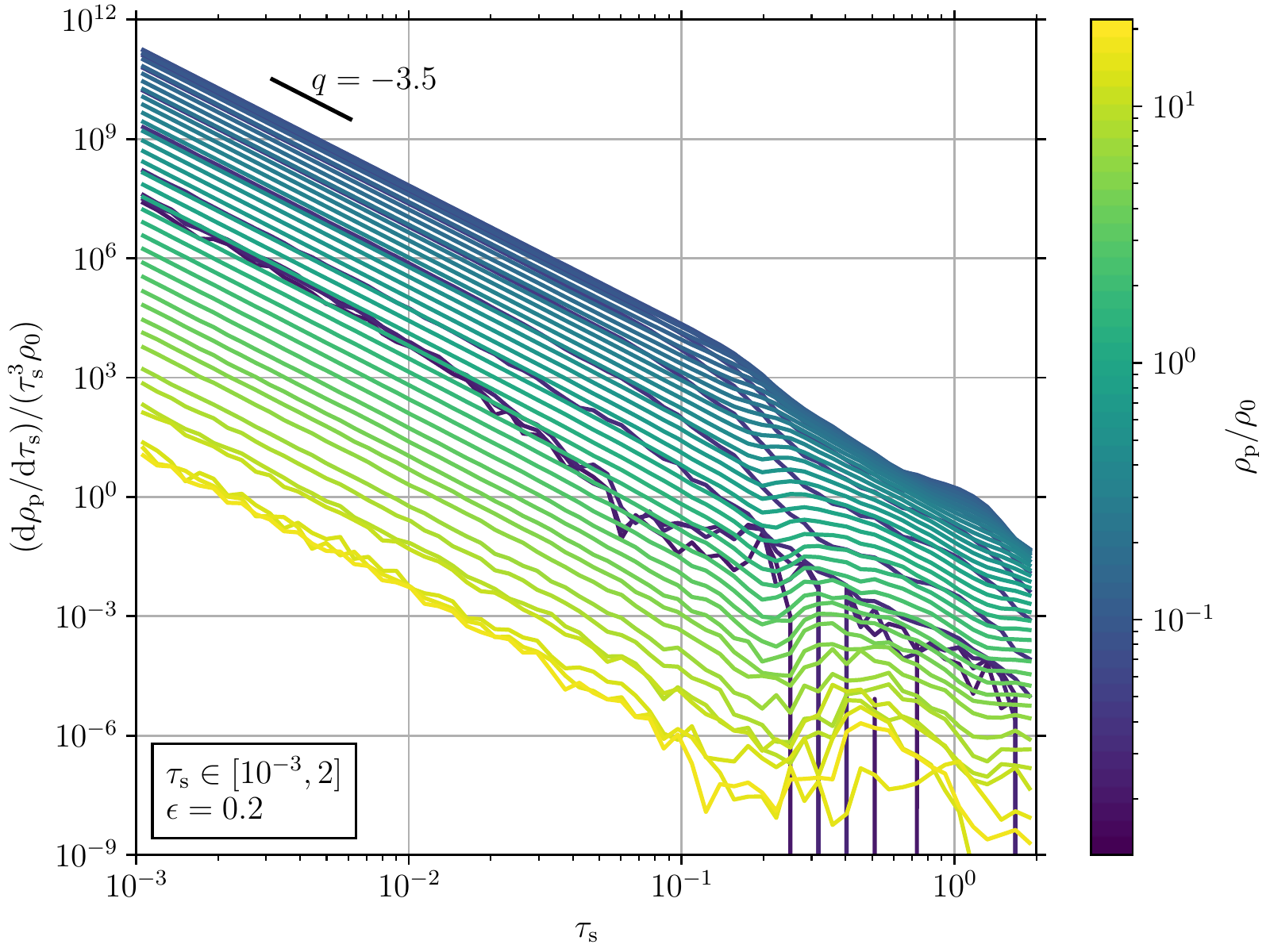}}
	\caption{Time-averaged dust-size distributions at varying total dust density $\rhop$ at the saturation state of each model with number of dust species $\Nsp = 64$.
	    The horizontal axis is the dimensionless stopping time $\taus$, which is approximately linear to the dust size $s$.
	    The vertical axis is a proxy to $\deriv{\np}{s}$, where $\diff{\np}$ is the number density of dust particles with sizes between $s$ and $s + \diff{s}$ (see Section~\ref{SS:sfd}).
	    Each line is the size distribution at different $\rhop$, and is coloured by its value as shown by the colour bar.
	    The black line segment indicates the slope of the input size distribution which has a power-law index $q = -3.5$.}
	\label{F:sfd}
\end{figure}

The dust segregation observed above implies that the dust-size distribution alter with location.
To demonstrate this, we find as follows the size frequency distribution (SFD) of dust particles located in all cells with a total dust density in between $\rhop$ and $\rhop + \diff{\rhop}$.
We begin with $\deriv{\rhop}{\taus}$, where $\diff{\rhop}$ is the sum of the masses of all particles with dimensionless stopping time in between $\taus$ and $\taus + \diff{\taus}$ in these cells then divided by the total volume of these cells.
Given that $\diff{\rhop} = m_\mathrm{p}\diff{\np} \propto s^3\diff{\np}$, where $m_\mathrm{p}$ is the mass of a dust particle, $\diff{\np}$ is the number density of particles with the same $\taus$, and $s$ is the size of the particles, the SFD is then
\begin{equation}
    \tder{\np}{s} \propto \frac{1}{s^3}\tder{\rhop}{s}
                  \propto \frac{1}{\taus^3}\tder{\rhop}{\taus},\label{E:sfd}
\end{equation}
where we have used the fact that $\taus$ is to a good approximation linear to $s$ \citep[see, e.g.,][]{sW77}.
Using equation~\eqref{E:sfd}, we plot in Fig.~\ref{F:sfd} the time-averaged SFD at the saturation state against the total dust density $\rhop$ for each of the three models above.
In other words, different lines in Fig.~\ref{F:sfd} represent the SFD observed in regions of different $\rhop$.
Each SFD can be compared with the power-law slope $q = -3.5$ of the input SFD (Section~\ref{S:method}).

As expected, the quiescent Model~As does not show noticeable change in the SFD with varying $\rhop$ (Fig.~\ref{F:sfd2}).
However, significant changes can be seen in the other two turbulent models.
For Model~Af, large particles are depleted in low-density regions and accumulate in high-density regions (Fig.~\ref{F:sfd1}).
Therefore, the SFD for $\taus \lesssim 0.01$ where $\rhop \lesssim \rho_0$ becomes much steeper than $q = -3.5$.
In high-density regions where $\rhop \gtrsim \rho_0$, the power-law slope remains $q \simeq -3.5$ up to $\taus \sim 0.01$, while the SFD levels off for $\taus \gtrsim 0.01$, indicating even heavier mass contributed by the largest particles.
For Model~B, similar effects can be seen but within a much smaller range of $\taus$ (Fig.~\ref{F:sfd3}).
In all regions, the slope of the SFD is maintained for $10^{-3} \leq \taus \lesssim 0.1$, while dust segregation is prominent for $\taus \gtrsim 0.1$, to the extent that even a bump is present in the SFDs in high-density regions.
We discuss potential implications of this dust segregation for observations in Section~\ref{S:cr}.

%-----------------------------------------------------------------------
\subsection{Radial drift and velocity dispersion} \label{SS:vp}

\begin{figure*}
	\centering
	\subcaptionbox{Model~Af\label{F:kin1}}
		{\includegraphics[width=0.33\textwidth]{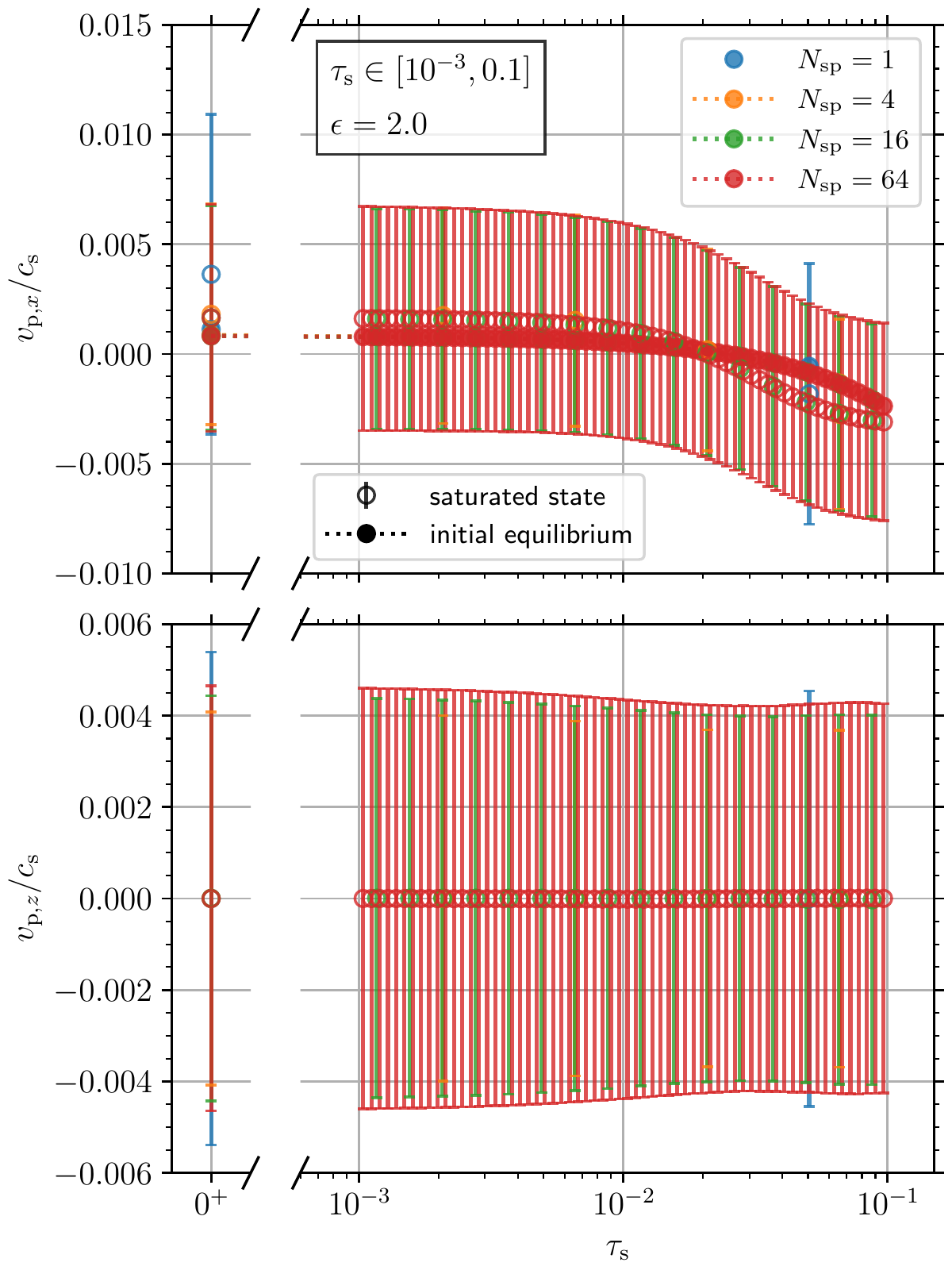}}
	\subcaptionbox{Model~As\label{F:kin2}}
		{\includegraphics[width=0.33\textwidth]{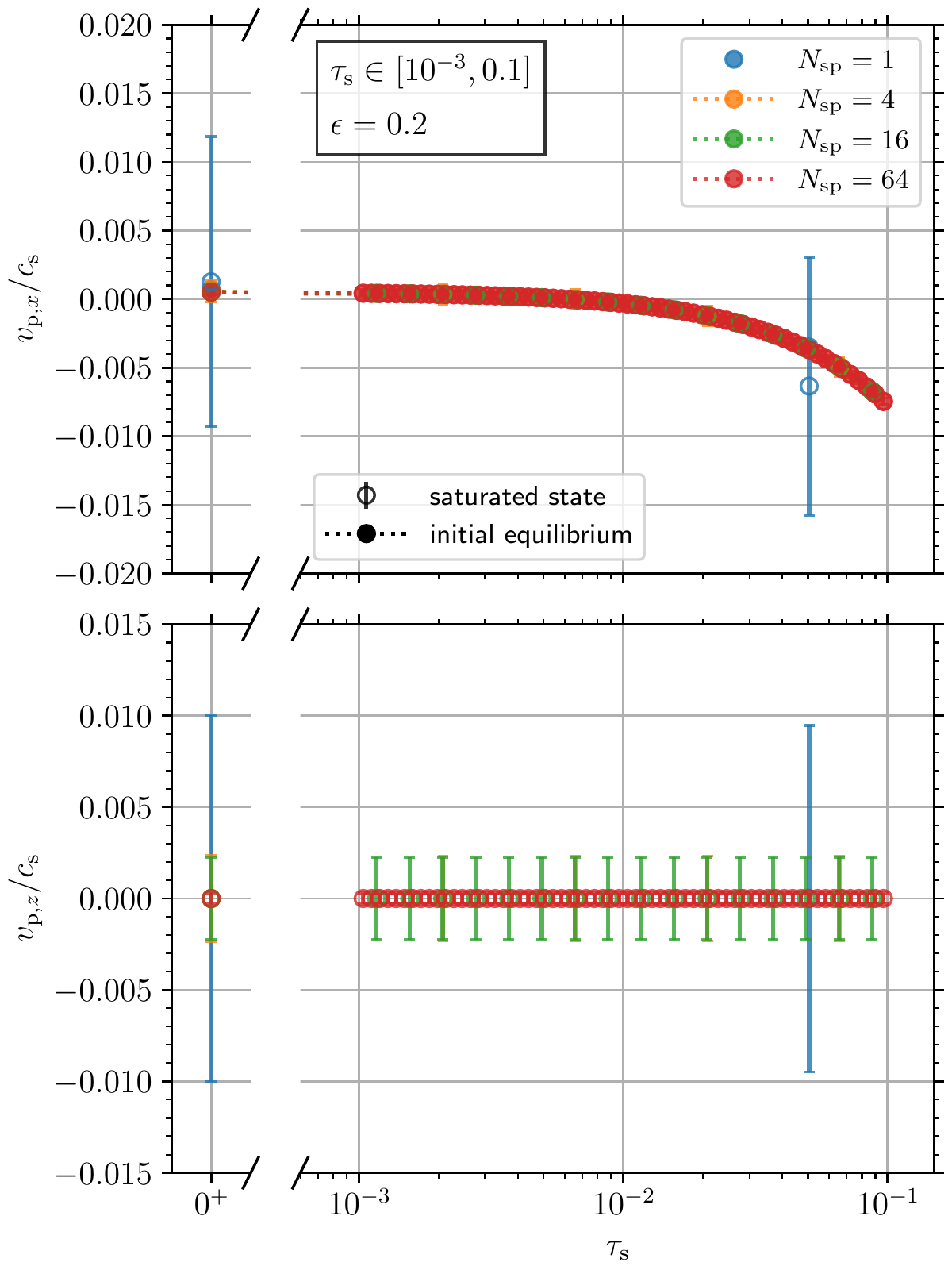}}
	\subcaptionbox{Model~B\label{F:kin3}}
		{\includegraphics[width=0.33\textwidth]{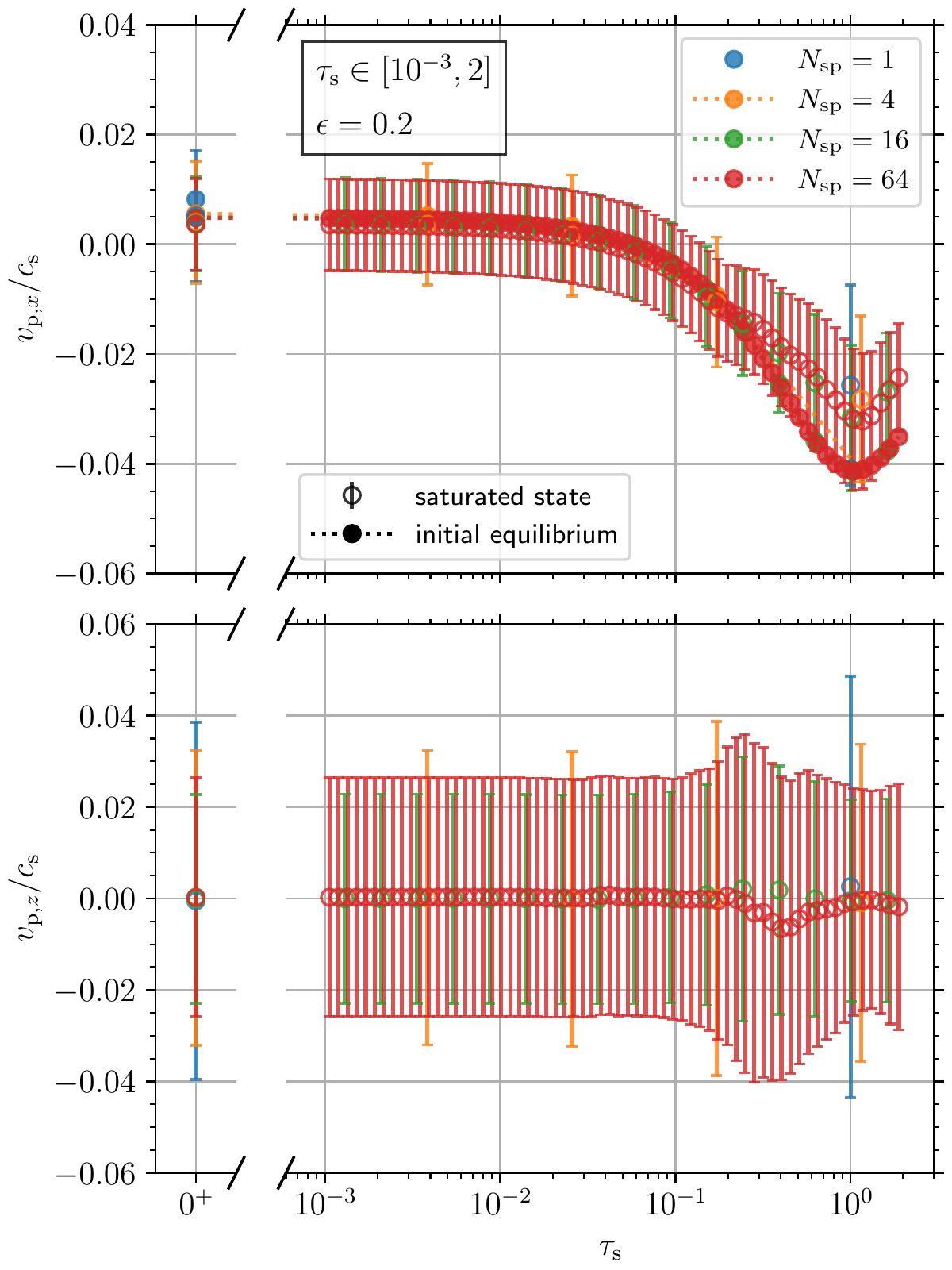}}
	\caption{Time-averaged mean velocity (\emph{open circles}) and velocity dispersion (\emph{vertical bars}) of each dust species as well as the gas (shown as $\taus \rightarrow 0^+$) at the saturation state.
	    The top panel shows the radial component, while the bottom panel shows the vertical component.
	    The mean radial velocity is compared with the initial equilibrium radial velocity (\emph{solid circles}).
	    Different colour indicates different number of species $\Nsp$ used to represent the dust-size distribution.
	    All velocities are normalised by the speed of sound $\cs$.}
	\label{F:kin}
\end{figure*}

Next, we analyse the kinematics of the dust particles.
We separate the particles by species, measure their mean velocity and velocity dispersion for each species as a function of time, and then take time averages at the saturation state.
The results are shown in Fig.~\ref{F:kin}, along with the same measurement for the gas taken from Table~\ref{T:turb}.
Over-plotted are the initial equilibrium radial velocities of the gas and the dust species, $u_{\mathrm{g}0,x}$ and $v_{\mathrm{p}0,j,x}$, respectively (Section~\ref{S:method}).

It is evident that the mean radial velocity deviates from the initial equilibrium velocity when the saturation state is turbulent, i.e., when the streaming instability is in the fast-growth regime.
Initially, the large species drift radially inwards while the small species and the gas drift outwards.
For Model~Af at the turbulent saturation stage, the dust species and the gas drift \emph{faster} on average than the initial state by a factor of a few, while largely maintaining their mean directions, except for few species with $\taus \simeq 0.02$--0.03 which are nearly stationary (Fig.~\ref{F:kin1}).
It appears that the dust-gas vortices generated in the system (Figs.~\ref{F:dens1} and~\ref{F:vort1}) are able to tap more energy into the separation of radial drifts between species.
For Model~B, by contrast, the radial drifts of the dust species and the gas exhibit an opposite trend (Fig.~\ref{F:kin3}).
The large species with $\taus \gtrsim 0.2$ and the small species with $\taus \lesssim 0.04$ as well as the gas drift \emph{slower} than the initial state by about 25\%.
From Fig.~\ref{F:dens3}, the collection of large dust species into dense filaments apparently reduces their own drift speeds by stronger back reaction to the gas.

The two cases also exhibit appreciable velocity dispersion in each dust species as well as in the gas.
For Model~Af (Fig.~\ref{F:kin1}), both the radial and vertical dispersions are relatively constant at $\sim$0.004--0.005$\cs$, with a slight trend of decreasing with increasing stopping time $\taus$.
For Model~B (Fig.~\ref{F:kin3}), the radial and the vertical dispersions of small species remain constant at $\sim$0.008$\cs$ and $\sim$0.026$\cs$, respectively, up to $\taus \simeq 0.1$.
The two components of the velocity dispersions for large species with $\taus \gtrsim 0.1$ are significantly excited, reaching $\simeq$0.013$\cs$ at $\taus \simeq 1$ and $\simeq$0.037$\cs$ at $\taus \simeq 0.3$, respectively.

In both cases, the kinematics of the dust particles appears to converge with the number of dust species $\Nsp$ representing the distribution.
The radial components of the mean velocity and velocity dispersion achieve consistent values between $\Nsp = 16$ and $\Nsp = 64$.
Although the vertical velocity dispersion does not reach consistent values between the two highest $\Nsp$'s, it can be seen that the difference decreases noticeably with increasing $\Nsp$.

Finally, as discussed in previous sections, Model~As is in the slow-growth regime and has a nearly quiescent saturation state.
As shown in Fig.~\ref{F:kin2}, the mean velocity at the saturation state hardly deviates from the initial equilibrium velocity, and each component of the velocity dispersion consistently decreases with increasing $\Nsp$, reaching $\sim$$3\times10^{-4}\cs$ at $\Nsp = 64$, in stark contrast with the saturation state of single species.

%-----------------------------------------------------------------------
\subsection{Turbulent diffusion} \label{SS:diff}

\begin{figure}
	\includegraphics[width=\columnwidth]{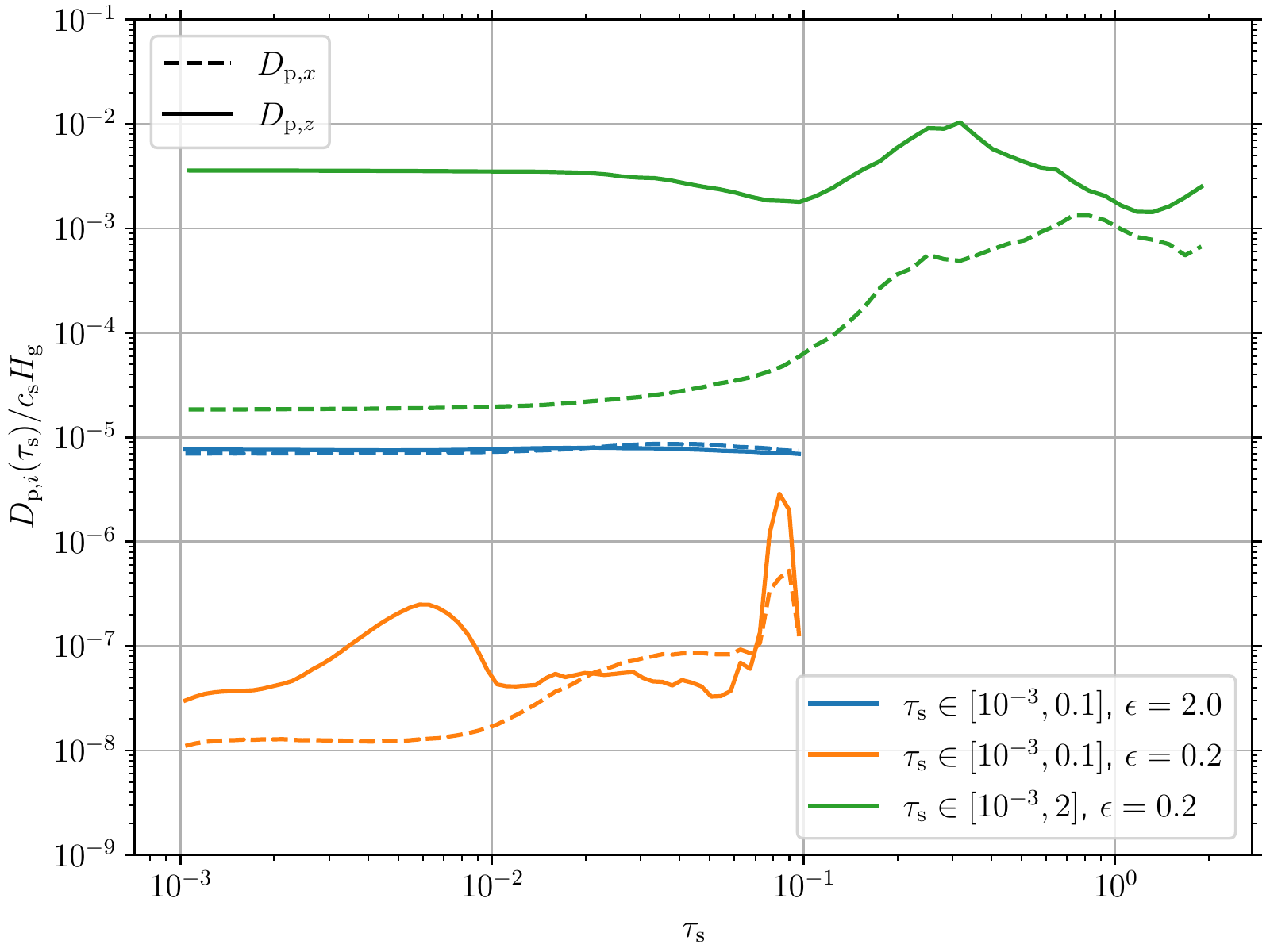}
	\caption{Time-averaged dust diffusion coefficients as a function of dimensionless stopping time $\taus$ at the saturation state of each model with number of species $\Nsp = 64$.
	The dashed lines show the radial diffusion coefficient $\Dpx$ while the solid lines the vertical one $\Dpz$.
	Different colours represent models with different dust-size distributions or solid-to-gas density ratios $\epsilon$.}
	\label{F:dcoeff}
\end{figure}

In the following, we investigate turbulent diffusion of dust particles at the saturation state.
We follow the procedure of \cite{YMM09} to measure the diffusion coefficient of each dust species in radial and vertical directions resulting from our models.
Starting at the saturation time listed in Table~\ref{T:turb}, the standard deviation of the displacement of all particles in the same species as a function of time is computed.
As the dependence of the standard deviation on the square root of time is recovered and the corresponding best fit is found, the coefficient of the best fit is then directly related with and converted into the diffusion coefficient $\Dpi{i}(\taus)$ with $i$ being $x$ or $z$.
The results for our models with number of species $\Nsp = 64$ are shown in Fig.~\ref{F:dcoeff}.

The blue lines in Fig.~\ref{F:dcoeff} represent Model~Af, the saturation state of which is driven by dust-gas vortices (Figs.~\ref{F:dens1} and~\ref{F:vort1}).
Both the radial and the vertical diffusion coefficients are appreciably close to a constant of magnitude 7--$9\times10^{-6}\cs\Hg$ over the entire range of stopping time.
Particles of larger sizes are slightly agitated, with $\Dpx$ and $\Dpz$ peaked at $\taus \simeq 0.03$ and 0.02, respectively.
We note that $\Dpi{x} \simeq \Dpi{z} \sim 10^{-5}\cs\Hg$ in our Model~Af is similar to what was found in previous vertically stratified simulations with $\tausmax = 0.1$ except for $\Dpi{z}$ of small particles (\citealt{BS10c,SYJ18}; see also Section~\ref{SS:hp} as well as \citealt{LY21}).

By contrast, the dust diffusion in Model~As is significantly weaker, as shown by the orange lines in Fig.~\ref{F:dcoeff}.
The magnitude of the diffusion coefficients is on the order of $10^{-8}$--$10^{-7}\cs\Hg$, except for several largest dust species.
The coefficients for the few largest species can reach about $10^{-6}\cs\Hg$, however given the level of fluctuations is so low, as noted in Section~\ref{S:sat}, we cannot exclude the nature of this effect as numerical and this effect may be responsible for the high-frequency oscillations in dispersions seen in Fig.~\ref{F:disp2}.
In any case, the general level and trend of our measured diffusion coefficients is consistent with the quiescence of the saturation state observed in previous sections (Fig.~\ref{F:dens2}).

For Model~B, the dust diffusion is considerably stronger than that in the other two models, as shown by the green lines in Fig.~\ref{F:dcoeff}.
The radial and the vertical diffusion coefficients for small dust particles with $\taus \lesssim 10^{-2}$ are nearly constant at $\sim$$2\times10^{-5}\cs\Hg$ and $\sim$$3\times10^{-3}\cs\Hg$, respectively, and hence dust transport in the vertical direction is much easier than in the radial direction.
For larger particles, the vertical diffusion coefficient shows some variation but generally is confined in between $10^{-3}\cs\Hg$ and $10^{-2}\cs\Hg$, with the maximum at $\taus \simeq 0.3$, while the radial diffusion coefficient shows monotonic increase with increasing $\taus$ until $\taus \sim 1$, where $\Dpx \sim 10^{-3}\cs\Hg$.
We note that in contrary to $\Dpi{z} > \Dpi{x}$ in our Model~B, previous vertically stratified simulations with $\tausmax = 1$ showed that $\Dpi{z} \lesssim \Dpi{x}$ (\citealt{BS10c,SYJ18}; see also \citealt{LY21}), and we discuss this further in Section~\ref{SS:hp}.

%-----------------------------------------------------------------------
\subsection{Implications for vertical scale height} \label{SS:hp}

\begin{figure}
	\includegraphics[width=\columnwidth]{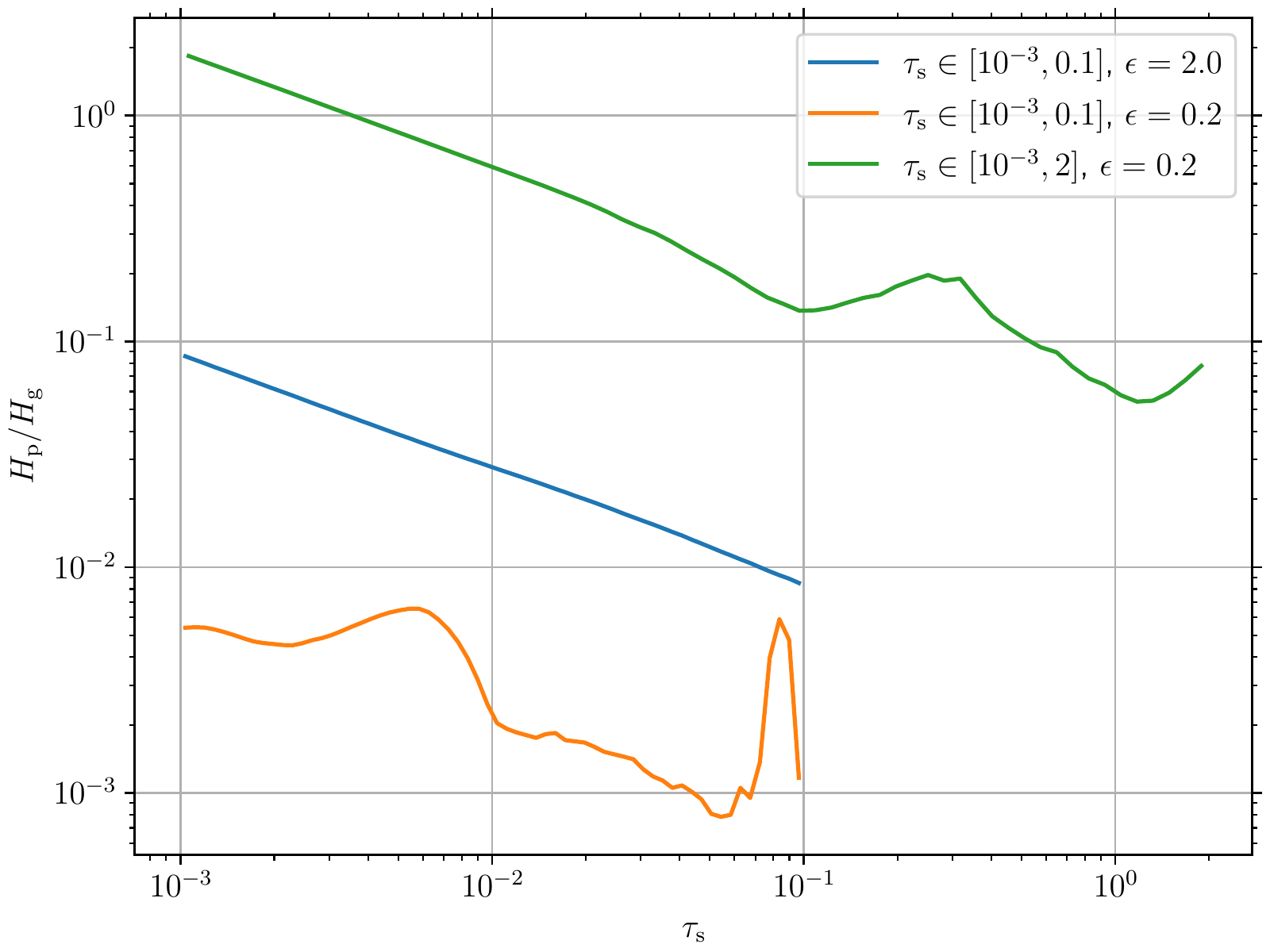}
	\caption{Estimated scale height of dust particles $\Hp$ as a function of dimensionless stopping time $\taus$ at the saturation state of each model with number of dust species $\Nsp = 64$, using equation~\eqref{E:hp}.
	The arrangement of colours representing different models is the same as in Fig.~\ref{F:dcoeff}.}
	\label{F:hp}
\end{figure}

With the vertical diffusion coefficient $\Dpz$, the scale height $\Hp$ of each dust species in the system can be estimated by equating the diffusion timescale $\Hp^2 / \Dpz$ with the sedimentation timescale $(\taus + \taus^{-1}) / \Omega_\mathrm{K}$ \citep{YL07} and hence
\begin{equation} \label{E:hp}
\Hp
\simeq \sqrt{\frac{\Dpz}{\Omega_\mathrm{K}}\left(\taus + \frac{1}{\taus}\right)}
= \Hg\sqrt{\frac{\Dpz}{\cs\Hg}\left(\taus + \frac{1}{\taus}\right)},
\end{equation}
where $\Omega_\mathrm{K}$ is the Keplerian angular frequency.
The estimate for each of our models with number of dust species $\Nsp = 64$ is shown in Fig.~\ref{F:hp}.
Since $\Dpz$ is nearly constant for Model~Af and most of $\taus \ll 1$, the dust scale height follows $\taus^{-1/2}$ closely, decreasing from about 0.09$\Hg$ at $\taus = 10^{-3}$ to about 0.008$\Hg$ at $\taus = 0.1$.
For the quiescent Model~As, the dust scale height shows some variation in between $\sim$$10^{-3}\Hg$ and $\sim$$10^{-2}\Hg$.
Hence, it appears that the nonlinear saturation of the streaming instability, even in a quiescent state, can maintain a minimum dust scale height of about $10^{-3}\Hg$.
We note also that some largest particles have similar scale height as their smallest counterparts.

Model~B is worth some discussion.
The scale height of the particles with $\taus \lesssim 0.1$ also follows $\taus^{-1/2}$, but the particles are significantly more excited than those in Model~Af.
The smallest particles even reach over one gas scale height.
We note that the particles in this system could travel freely in the vertical direction between the dense filaments, and the typical separation between adjacent filaments is on the order of $\Hg$ (see Fig.~\ref{F:dens3}).
Therefore, the vertical mixing length is more than $O(\Hg)$, which may not be realised when the vertical component of the stellar gravity is included.
In other words, our estimate of the particle scale height for the dust distribution of Model~B is likely to be an upper limit.
On the other hand, it could be seen that in vertically stratified simulations, the dust scale height is in general larger for the distribution $\taus \in [10^{-3}, 1]$ than for $\taus \in [10^{-4}, 0.1]$ when compared at the same $\taus$ and the same solid abundance \citep{BS10c,SYJ18}, and hence we expect that a dust distribution with largest particles of $\taus \gtrsim 1$ indeed leads to a more vertically excited dust layer.
We note that though, the dust layer may still be thinner for the former distribution when comparing the leading sizes.
Moreover, the dust scale height in general reduces with increasing solid abundance in stratified simulations \citep{BS10c,YJC17,YMJ18}.
It will be of interest to investigate the morphological and kinematical structures of the particle layer near the mid-plane of vertically stratified simulations with  a dust-size distribution led by $\tausmax \gtrsim 1$, as compared to this work.

Moreover, Fig.~\ref{F:hp} brings into question the domain size of vertically stratified simulations with multiple dust species.
For our Model~Af, $\Hp \sim 0.1\Hg$ for $\taus \sim 10^{-3}$, while the majority of the previous works considered a domain size of $\sim$0.2$\Hg$ ($|z| \lesssim 0.1\Hg$).
We note that our estimate of the dust scale height from this model is similar with the measurement from previous vertically stratified simulations for $\taus \gtrsim 10^{-2}$ (Model~R41Z1 of \citealt{BS10c} and Model~SI41-4-4-d of \citealt{SYJ18}, both of which had the same $\tausmax = 0.1$).
On the other hand, \cite{SYJ18} showed increasing dust scale height for $\taus \lesssim 10^{-2}$ with increasing domain size up to 0.8$\Hg$, and no convergence has been found for these small particles.
Hence, future systematic investigation of vertically stratified simulations with multiple dust species on domain size seems warranted.

%-----------------------------------------------------------------------
\subsection{Implications for radial transport and mixing} \label{SS:rtm}

\begin{figure}
	\includegraphics[width=\columnwidth]{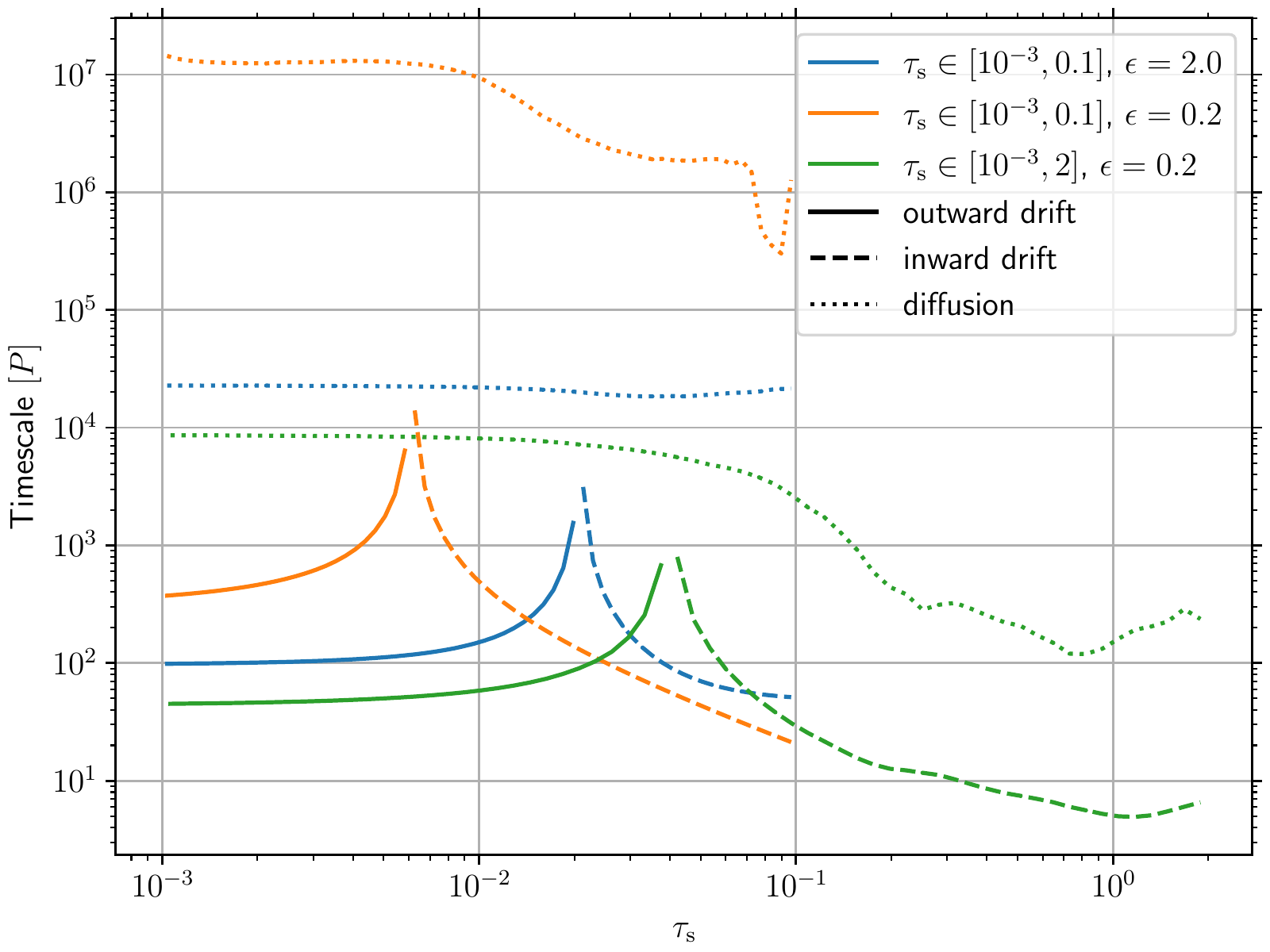}
	\caption{Timescales in terms of orbital period $P$ for radially transporting dust particles over one gas scale height $\Hg$ as a function of dimensionless stopping time $\taus$.
	The transport processes include outward drift (\emph{solid} lines), inward drift (\emph{dashed} lines), and diffusion (\emph{dotted} lines).
	The arrangement of colours representing different models is the same as in Fig.~\ref{F:dcoeff}.}
	\label{F:ts}
\end{figure}

The radial transport of dust particles consists of both radial drift and turbulent diffusion.
We compare their relative importance by computing the timescale for each process to transport a dust species over one gas scale height $\Hg$.
The radial drift timescale is given by $\Hg / \left|\overline{v_{\mathrm{p},j,x}}\right|$, where $\overline{v_{\mathrm{p},j,x}}$ is the mean radial velocity of the $j$-th species at the saturation state measured in Section~\ref{SS:vp}.
The diffusion timescale is given by $\Hg^2 / \Dpx$, where $\Dpx$ is the radial diffusion coefficient measured in Section~\ref{SS:diff}.
The results for our models with the number of species $\Nsp = 64$ are shown in Fig.~\ref{F:ts}.

It can be seen that radial drift dominates over turbulent diffusion in all cases (see also \citealt{BS10c} and \citealt{SYJ18}), except for those dust species in a small range of sizes that have nearly zero radial drift.
The two timescales differ by at least two orders of magnitude, and the difference further increases with larger distance scales.
We note also that at the saturation state of the streaming instability in the fast-growth regime (Models~Af and~B), the outward drift of small particles is not significantly slower than the inward drift of large particles, in contrast to what the initial equilibrium (Section~\ref{SS:vp} and Fig.~\ref{F:kin}) would indicate if turbulence were not developed.
This may have potential consequences on the radial mixing of dust materials in protoplanetary discs, as further discussed in Section~\ref{S:cr}.

%-----------------------------------------------------------------------
\section{Concluding Remarks} \label{S:cr}

In this work, we build upon the findings in the linear analysis of multi-species streaming instability in Paper~I and continue to investigate the nonlinear saturation of the instability using numerical simulations.
We focus on three distinct combinations of the dust-size distribution and the total solid-to-gas density ratio $\epsilon$ (Table~\ref{T:specs}).
Two of them are in the fast-growth regime: one has a high $\epsilon = 2$ but a low maximum dimensionless stopping time $\tausmax = 0.1$ (Model~Af), and the other has a high $\tausmax = 2$ but a low $\epsilon = 0.2$ (Model~B).
The third is in the slow-growth regime and has both a low $\tausmax = 0.1$ and a low $\epsilon = 0.2$ (Model~As).
The minimum $\taus$ is fixed at $\tausmin = 10^{-3}$.
We systematically vary the number of discrete dust species up to $\Nsp = 64$ that represents the dust-size distribution (as well as the resolution; Appendix~\ref{S:res}), and study the properties of the saturation state of the models and their convergence.

For the two cases in the fast-growth regime, we find that the dust-gas dynamics at the saturation state is qualitatively similar to their counterparts in the single-species streaming instability with a dust size similar to the largest sizes in the distribution of the former.
When \dda{} with $\epsilon = 2$, the larger particles undergo vortical motions along with the gas and collect in between vortices (Fig.~\ref{F:dens1}).
On the other hand, when \ddb{} with $\epsilon = 0.2$, the larger particles undergo radial traffic jams and collect in dense filamentary structures (Fig.~\ref{F:dens3}).
We note that, however, the maximum dust density reached in this case is significantly lower than the single-species counterpart (Fig.~\ref{F:rpdf3}).
Furthermore, the dust density distribution for the case of \dda{} with $\epsilon = 2$ is flat in low-density regions (Fig.~\ref{F:rpdf1}), while the case of \ddb{} with $\epsilon = 0.2$ has a diffuse background of small particles and hence its density distribution shows a sharp low-end cutoff (Fig.~\ref{F:rpdf3}).
In any case, the smaller particles remain relatively diffuse and the larger particles tend to concentrate, resulting in noticeable dust segregation in sizes (Section~\ref{SS:sfd} and Fig.~\ref{F:sfd}).
Finally, we find that $\Nsp \gtrsim 16$ should be sufficient to obtain consistent turbulence properties in both cases.

By contrast, the saturation state of the case in the slow-growth regime (\dda{} with $\epsilon = 0.2$) appears to be quiescent (Fig.~\ref{F:dens2}) as compared with the previous two cases.
All the diagnostics, including gas turbulence (Table~\ref{T:turb}), width of the dust density distribution (Fig.~\ref{F:rpdf2}), dust velocity dispersion (Fig.~\ref{F:kin2}), and turbulent diffusion of dust particles (Fig.~\ref{F:dcoeff}), monotonically decreases with increasing $\Nsp$, and no evident convergence to a finite level can be seen up to $\Nsp = 64$.
Nevertheless, this state may still maintain a vertical scale height of the dust layer at $\Hp \gtrsim 10^{-3}\Hg$ across the range of the dust sizes, at least when $\Nsp = 64$ (Fig.~\ref{F:hp}).

As discussed above and in Section~\ref{SS:sfd}, dust particles in streaming turbulence may segregate by size, and this may have consequences in observations by Atacama Large Millimeter/submillimeter Array (ALMA), specifically on the interpretation of multi-wavelength spectral index \citep[e.g.,][]{CS19}.
Figure~\ref{F:sfd} shows that the power-law slope of the dust-size distribution may deviate from that of the overall distribution, depending on the local total dust density.
If the densest region in dust concentration driven by the streaming instability remains optically thin, the observations may not discern this subtle effect of dust segregation.
However, should the dense regions become optically thick, the observed spectral index can be skewed toward a dust-size distribution in diffuse regions, where large particles of some range of sizes are depleted.
In this case, the spectral index may either be steeper or flatter than the overall slope, depending on which part of the dust-size distribution an observation is sensitive to.

An understanding of how solid materials from $\micron$-sized dust to mm/cm-sized pebbles transport and mix is essential to reconstruct the history of the early Solar nebula from the meteoritic records and the sample returns of Solar System bodies \citep[see, e.g.,][]{DA14,KN15}.
As discussed in Section~\ref{SS:rtm}, turbulence driven by the streaming instability may play a role in two folds.
First, streaming turbulence may alter the radial drift speeds of solid particles of different sizes (Fig.~\ref{F:kin}).
For the fast-growth regime, the timescales of outward and inward migrations can become comparable, making the former potentially as significant as the latter in contrast to what would be predicted by the drag-force equilibrium (Fig.~\ref{F:ts}).
Second, even though diffusion of dust particles in streaming turbulence is not as dominant as differential radial drift in enhancing radial mixing (Fig.~\ref{F:ts}; see also \citealt{BS10c,SYJ18}), the timescale of the former remains short, especially in the inner disc of a few au, with respect to the typical lifetime of a protoplanetary disc of a few Myr \citep[e.g.,][]{WC11}.
In addition, the importance of turbulent diffusion increases with decreasing length scales, especially when $\lesssim$0.01--0.1$\Hg$, in which materials could be homogenised.
Therefore, both mean radial drift and radial diffusion of dust particles according to their sizes should be considered in a model of the early Solar nebula as well as of protoplanetary discs.

Last but not least, the findings in this work may have several potential implications for planetesimal formation.
First of all, as suggested by Fig.~\ref{F:rpdf3} and discussed above, the maximum dust density reached in the saturation state of the multi-species streaming instability with the largest sizes of $\tausmax \gtrsim 1$ is significantly smaller than its single-species counterpart.
Without vertical sedimentation, the traffic jams in this unstratified disc may not be sufficient to drive local dust concentrations over the Roche density and trigger gravitational collapse \cite[see also][]{YJC17}.
Second, the dust segregation discussed in Section~\ref{SS:sfd} indicates that significantly more large particles are present in dense regions of solids than on average (Fig.~\ref{F:sfd}).
Since planetesimals are predisposed to form from these regions, this implies that the initial dust-size distribution in a pebble cloud before collapse may be skewed towards large sizes from the background distribution, which may have some consequences in the process of gravitational collapse and the final composition and radial structure of a newborn planetesimal \citep{WJ17,PM21,VDD21}.
Third, as found in Section~\ref{SS:vp}, the vertical velocity dispersions of dust particles are similar to or significantly larger than the radial ones (Model~Af and Model~B, respectively; Fig.~\ref{F:kin}).
This implies that planetesimals can form with a wide range of obliquity, which would be similar to the findings by \cite{NL19}.
Finally, in a vertically stratified system, the solid-to-gas ratio of \emph{column} densities $Z$ is related with that of volume densities in the mid-plane by
\begin{align}
    Z &\equiv \frac{\Sigma_\mathrm{p}}{\Sigma_\mathrm{g}}
       = \int\frac{\Hp(\taus)}{\Hg}\,\frac{\diff{\rhop(\taus)}}{\rhog}\nonumber\\
      &= \frac{(4+q)\epsilon}{\tausmax^{4+q} - \tausmin^{4+q}}
         \int_{\tausmin}^{\tausmax}
         \frac{\Hp(\taus)}{\Hg}\taus^{4+q}\,\diff{\ln\taus},
\end{align}
where $q$ is the power-law index of the dust-size distribution (Section~\ref{S:method}).
Using this formula along with Fig.~\ref{F:hp}, the model of \dda{} with $\epsilon = 0.2$, a quiescent state, implies that $Z \approx 0.05\%$.
If the turbulent diffusion of dust particles continues to weaken with increasing $\Nsp$, $\Hp / \Hg$ and hence our estimate of $Z$ may also be lowered.
On the other hand, the model of the same distribution but high $\epsilon = 2$, a turbulent state, implies that $Z \approx 4\%$.
These two estimates coincidentally straddle the critical $Z$ observed in numerical simulations with vertical gravity and multiple dust species \citep{BS10c,SJL21}.
Therefore, the sharp boundary of $\epsilon \sim 1$ for $\tausmax \lesssim 1$ separating the fast- and slow-growth regimes found in Paper~I might have some interesting ramifications to the vertically stratified system and the resulting conditions for planetesimal formation.
More studies are required to unravel the potential connection between the turbulence driven by the streaming instability and the strong clumping of solids under vertical sedimentation.

%-----------------------------------------------------------------------
\section*{Acknowledgements}

We appreciate all of the detailed and useful comments made by our reviewer.
We would also like to thank Francesco Lovascio, Colin McNally, Sijme-Jan Paardekooper, and Urs Sch\"{a}fer for their comments on this work.
We are especially grateful for the linear growth rates in the continuum limit provided by Paardekooper for our models.
Resources supporting this work were provided by the NASA High-End Computing (HEC) Program through the NASA Advanced Supercomputing (NAS) Division at Ames Research Center.
This work also used the Extreme Science and Engineering Discovery Environment (XSEDE) Stampede2 at the Texas Advanced Computing Center (TACC) through allocation AST130002.
We are grateful for the support from NASA via the Emerging Worlds program (Grant Number 80NSSC20K0347) and via the Astrophysics Theory Program (Grant Number 80NSSC21K0141).
CCY is also grateful for the support from NASA via the Theoretical and Computational Astrophysics Networks program (Grant Number 80NSSC21K0497).
ZZ acknowledges the support from the National Science Foundation under CAREER Grant Number AST1753168.

%-----------------------------------------------------------------------
\section*{Data Availability}

The data underlying this article will be shared on reasonable request to the corresponding author.

%-----------------------------------------------------------------------
\bibliographystyle{mnras}
\bibliography{ms}

\appendix
\renewcommand{\thefigure}{A\arabic{figure}}
%-----------------------------------------------------------------------
\begin{figure*}
	\centering
	\subcaptionbox{Model~Af\label{F:rspdf1}}
		{\includegraphics[width=0.33\textwidth]{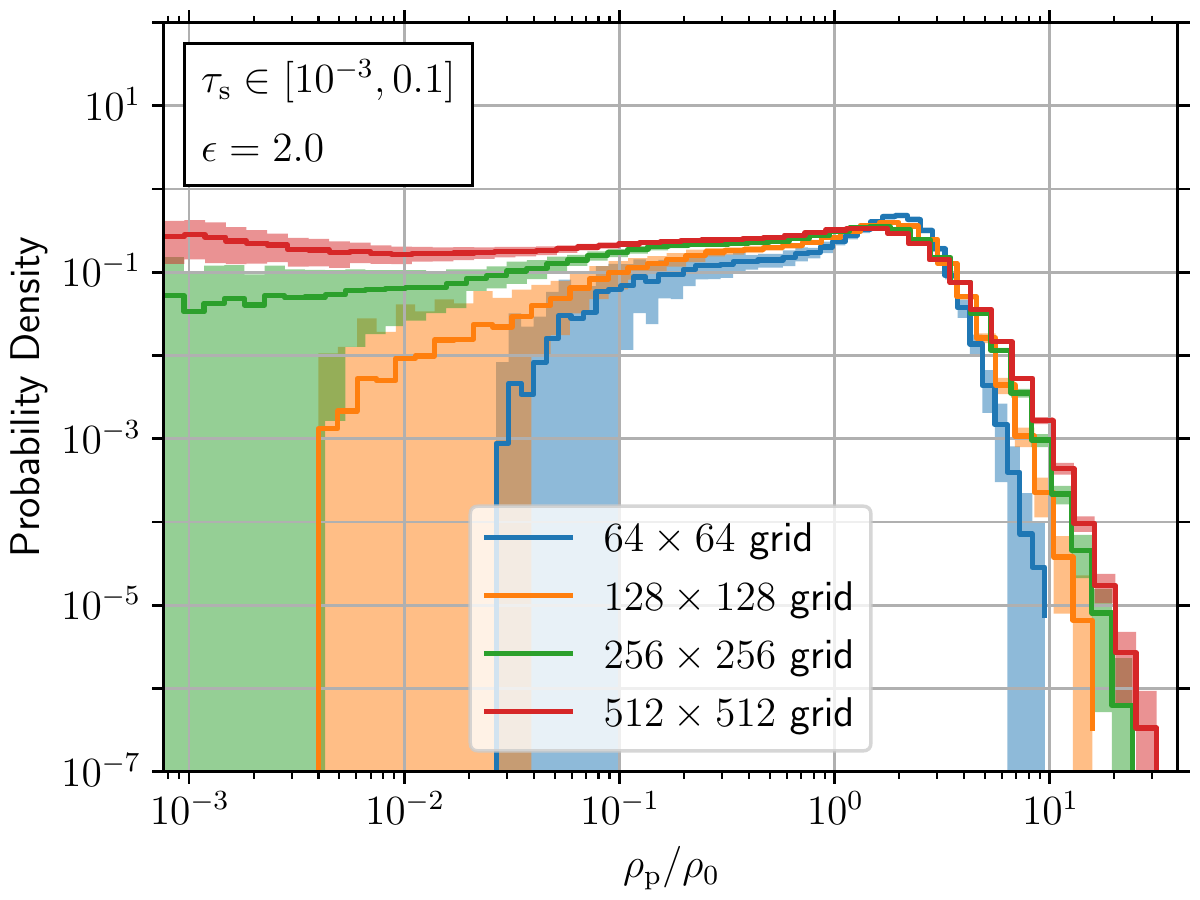}}
	\subcaptionbox{Model~As\label{F:rspdf2}}
		{\includegraphics[width=0.33\textwidth]{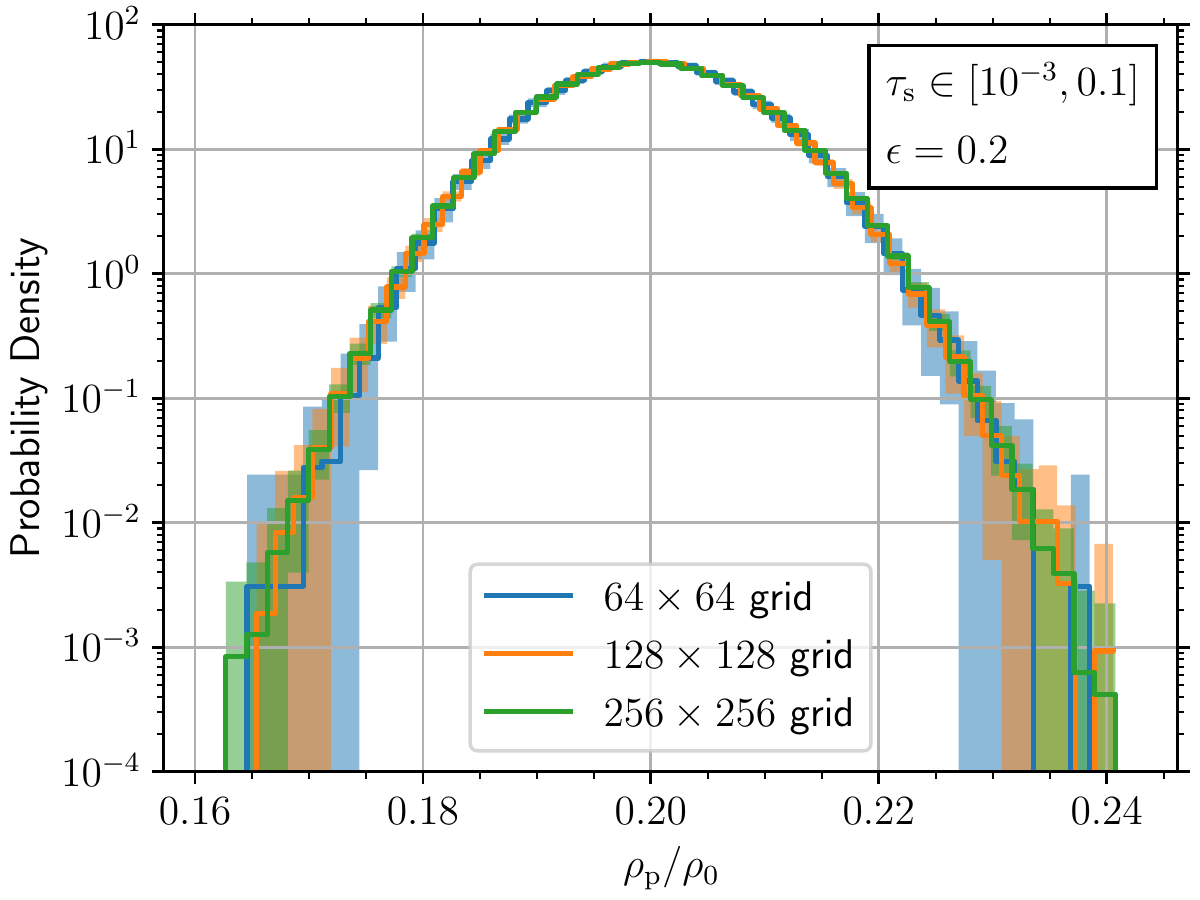}}
	\subcaptionbox{Model~B\label{F:rspdf3}}
		{\includegraphics[width=0.33\textwidth]{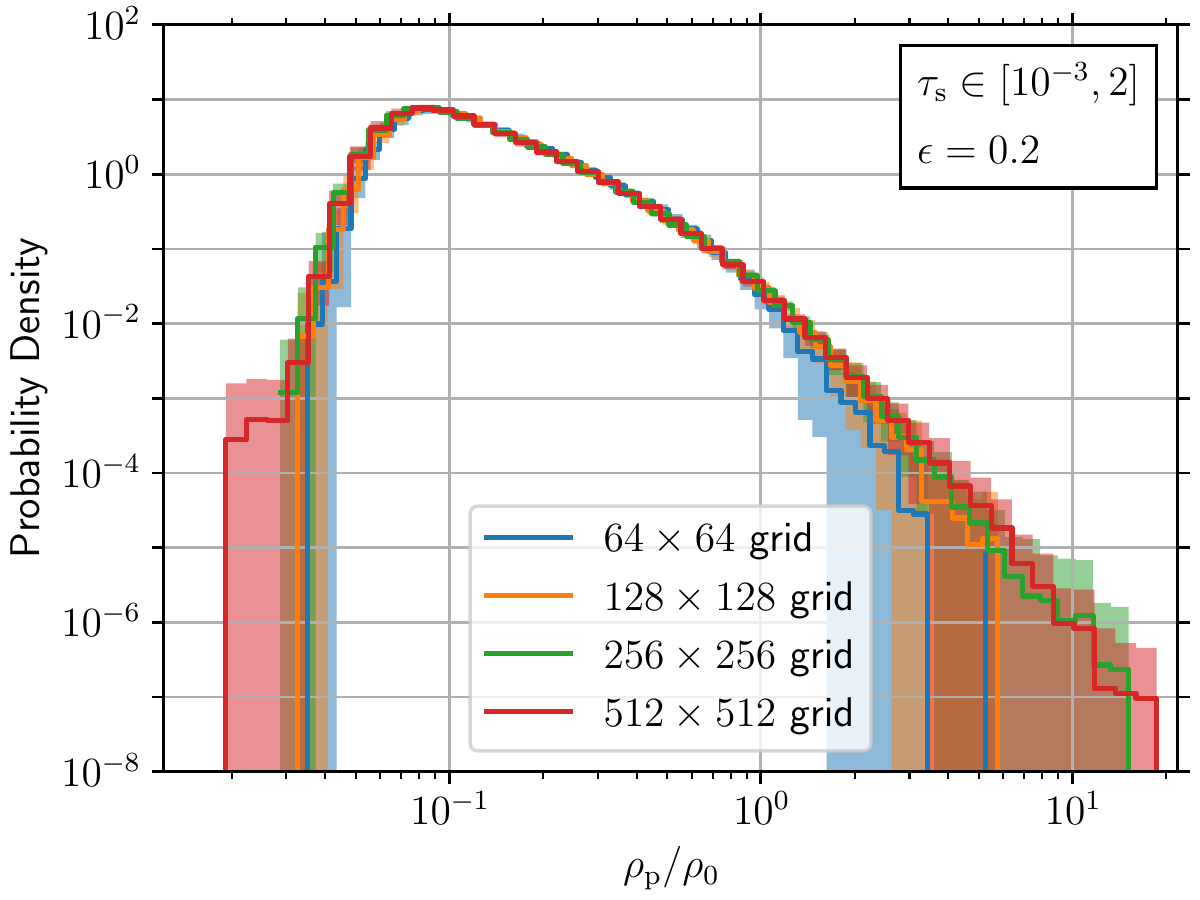}}
	\caption{Resolution study of time-averaged probability density as a function of the total dust density $\rhop$ for models with number of discrete dust species $\Nsp = 64$.
	    Each solid line shows the probability density, and the corresponding shade indicates the time variability.
	    Different colour represents different resolution.}
	\label{F:rspdf}
\end{figure*}

\begin{figure*}
	\centering
	\subcaptionbox{Model~Af\label{F:rskin1}}
		{\includegraphics[width=0.33\textwidth]{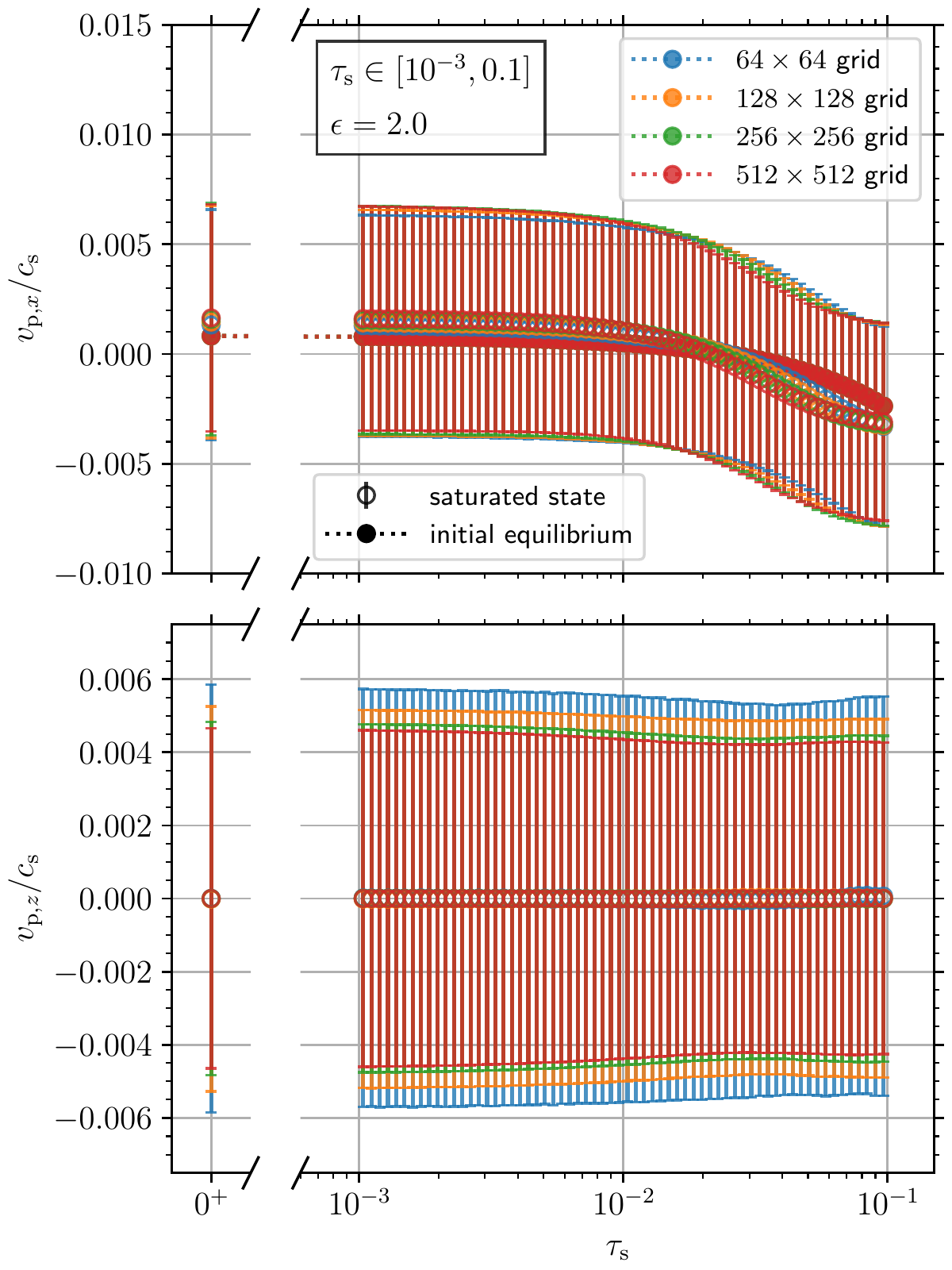}}
	\subcaptionbox{Model~As\label{F:rskin2}}
		{\includegraphics[width=0.33\textwidth]{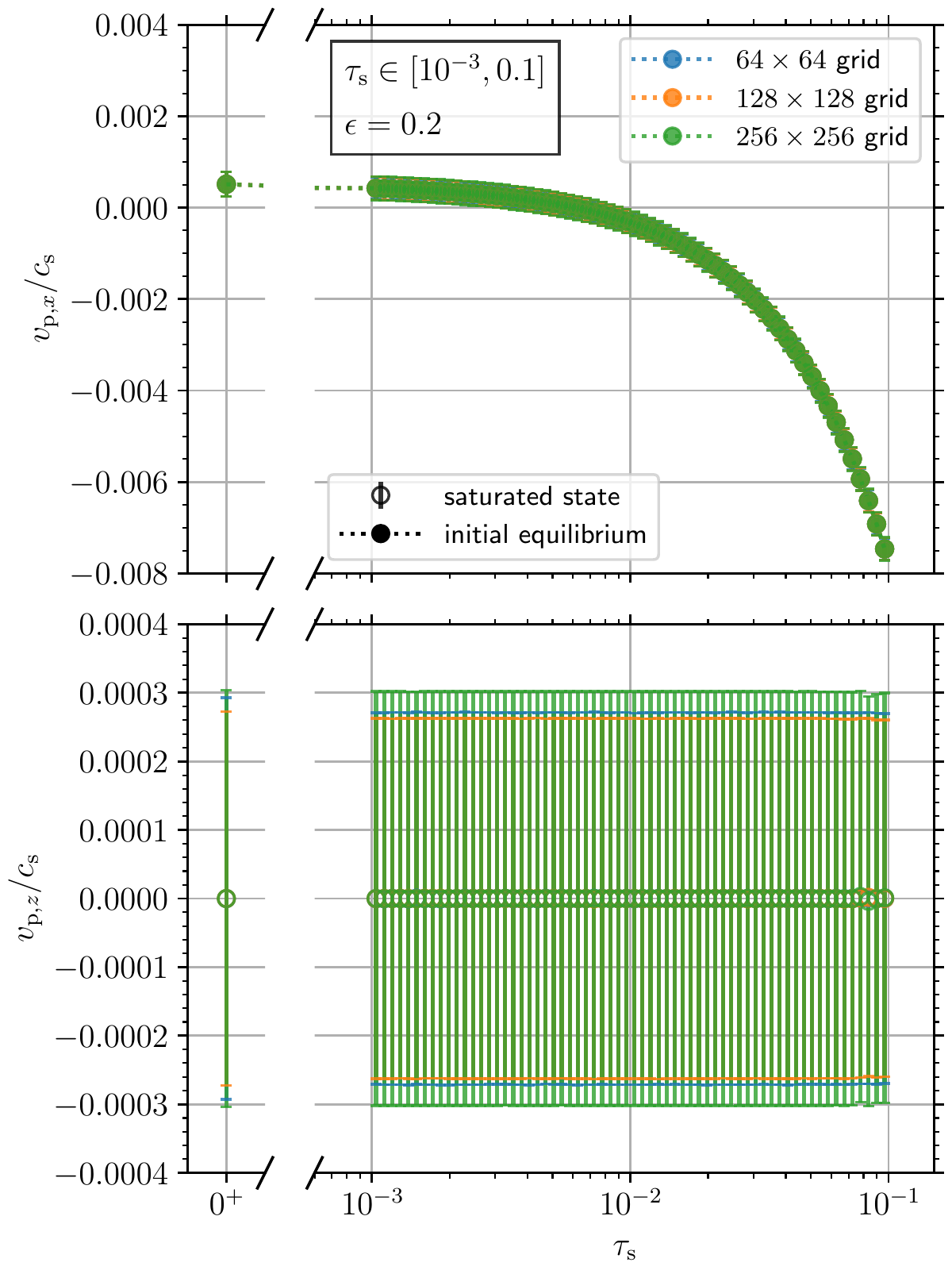}}
	\subcaptionbox{Model~B\label{F:rskin3}}
		{\includegraphics[width=0.33\textwidth]{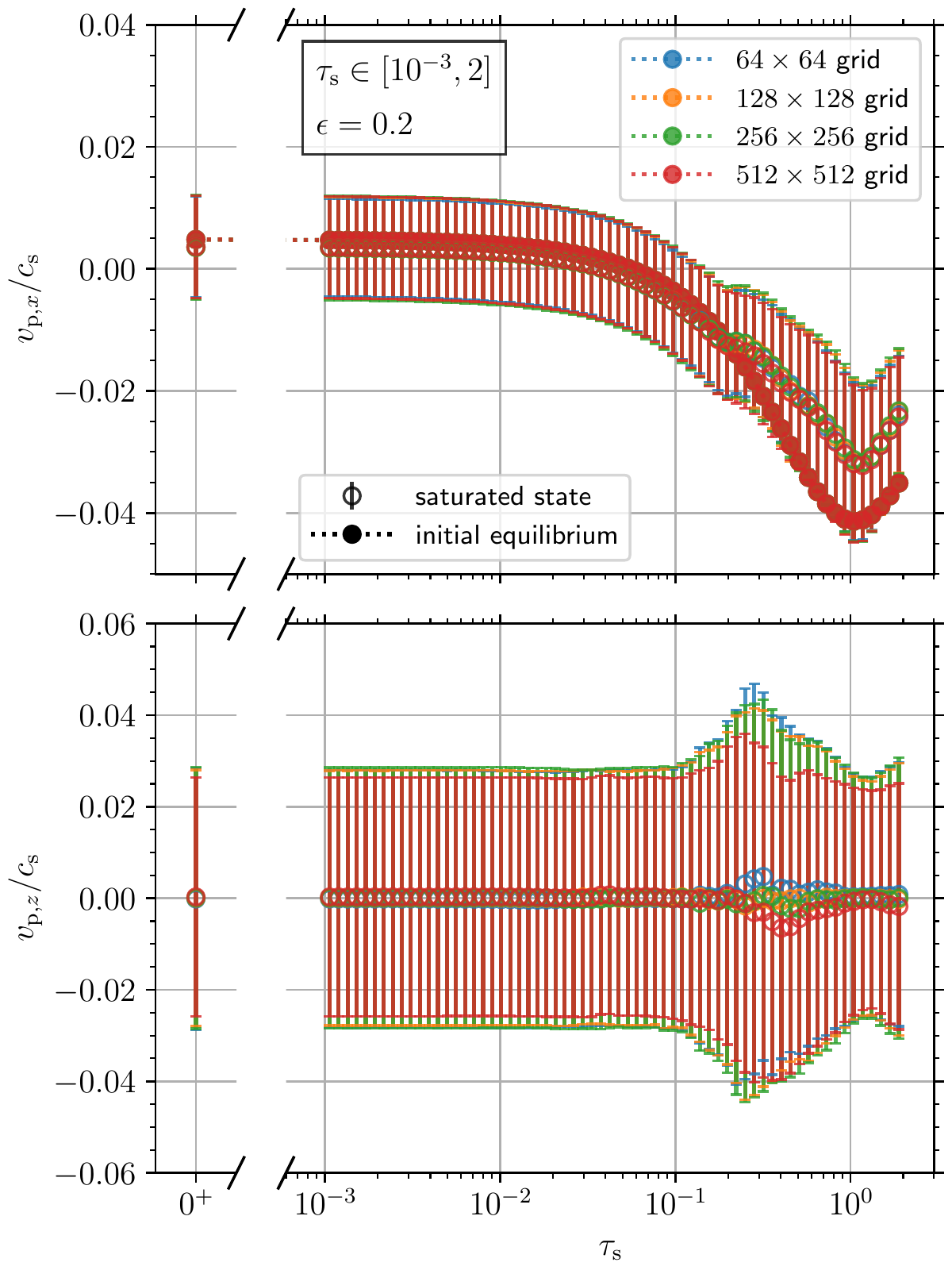}}
	\caption{Resolution study of time-averaged mean velocity (\emph{open circles}) and velocity dispersion (\emph{vertical bars}) of each dust species as well as the gas (shown as $\taus \rightarrow 0^+$) at the saturation state.
	    The top and the bottom panels show the radial and the vertical components, respectively.
	    Also plotted in the top panel is the initial equilibrium radial velocity (\emph{solid circles}).
	    All models have $\Nsp = 64$ discrete dust species that represent the dust-size distribution, and different colour represents different resolution.}
	\label{F:rskin}
\end{figure*}

\section{Resolution Study} \label{S:res}

For each combination of dust-size distribution and total solid-to-gas density ratio listed in Table~\ref{T:specs}, we have conducted resolution studies from a 64$\times$64 grid up to the maximum resolution for each of $\Nsp = 1$, 4, 16, and 64.
In this section, we show such studies with the distribution function of dust density (see Section~\ref{SS:dd}) and the dust kinematics (see Section~\ref{SS:vp}) in Fig.~\ref{F:rspdf} and Fig.~\ref{F:rskin}, respectively, for our models with $\Nsp = 64$.
In general, our models demonstrate satisfactory convergence with resolution in that either a variable has similar values at different resolutions or the difference of the values between pairs of resolutions decreases with increasing resolution.

A potential uncertainty may be the vertical velocities of large particles ($\taus \gtrsim 0.1$) for Model~B, as shown in the bottom panel of Fig.~\ref{F:rskin3}.
These particles are particularly excited in the vertical direction, and the momentum exchange appears random, leading to appreciable mean movement in either direction for different species, irrespective of the resolution.

%=======================================================================
% Don't change these lines
\bsp	% typesetting comment
\label{lastpage}
\end{document}